**Title:** Bigger aircraft, fewer emissions? Enumerating the technological viability and climate impact of jet electrification

**Authors:** Megan Yeo[1], Sebastian Nosenzo[2], Sichen Shawn Chao[3], and Ashley Nunes[1,4,5]

[1] Department of Economics
Harvard College
Cambridge, MA, 02138, USA

[2] Graded - The American School of São Paulo
São Paulo - SP, 05642-001, Brazil

[3] Sloan School of Management
Massachusetts Institute of Technology
Cambridge, MA, 02139, USA

[4] Center for Labor and a Just Economy
Harvard Law School
Cambridge, MA, 02138, USA

[5] Department of Urban Studies and Planning
Massachusetts Institute of Technology
Cambridge, MA, 02139, USA

**Corresponding Author:** Ashley Nunes, anunes@fas.harvard.edu

**Keywords**: electric aviation, climate change, grid decarbonization

**Abstract**

Enabling battery technology has not achieved sufficient maturity to facilitate electric flight for all aircraft models across all distances. Consequently, existing discourse emphasizes electrifying short haul routes using smaller, lighter aircraft. Does this emphasis have merit? Leveraging data for 47 different aircraft models, over 33 million commercial flights, and grid carbon intensity for 105 countries, we estimate a model that addresses these questions. Our findings are four-fold. First, we find that current energy density limitations impede short haul electric flight, regardless of aircraft model utilized. Second, we document that electrifying smaller, lighter aircraft models serving short haul routes may be particularly challenging as these aircraft require more (not less) acute increases in energy density (compared to larger, heavier aircraft models serving the same routes). Third, we identify a subset of larger, heavier aircraft as better candidates for electrification and note that doing so could prevent the annual release of at least 917,826,722 kilograms of carbon dioxide equivalent. However, we observe that the regional benefits of electrification are highly heterogeneous. The largest emissions benefit is realized in Europe, followed by South America, North America, Oceania and Africa. Electrification flights originating in Asia produces a net increase in carbon emissions owing to the disproportionate share of miles claimed by Asian countries with a more carbon intensive electrical grid. Three Asian countries - India, Saudi Arabia and Malaysia – emerge as top polluters, accounting for 37 percent of continent-wide miles but responsible for 67 percent of continent-wide emissions. India’s emissions impact warrants particular scrutiny, as its emissions contribution most disproportionately exceeds its mileage contribution. The implications of these findings for decarbonization policy are subsequently discussed.

## Introduction

Can electrification facilitate emissions reductions in the aviation sector? If so, to what extent? In 2019, nearly 4.5 billion passengers travelled by air, up from 100 million in 1960 (1,2). Although this increase – facilitated by deregulation and technological improvements – has been economically beneficial, externalities persist (2). Commercial aviation depends almost entirely on kerosene, a liquified hydrocarbon that while providing the large amount of energy needed for planes to get, and stay, airborne, also makes air travel among the most carbon-intensive form of transportation (3,4). Emissions estimates reflect these concerns. Annual carbon dioxide ($CO_2$) emissions from commercial aviation are currently estimated to be 186 million metric tons $CO_2$ and are projected to grow to 209 $CO_2$ by 2050 (5).

Given the associated climate consequences, stakeholders have intensified scrutiny of and investment in technologies that reduce aviation sector emissions (6-10). Electric-powered aircraft are one such technology. By some estimates, electric aircraft offer, owing to their reduced dependence on fossil fuels, a reduction in $CO_2$e emissions (relative to fossil-fueled reference aircraft) of up to 88 percent (10). This favorability persists even after the emissions intensity of airframe and battery manufacturing is accounted for. Consequently, policymakers worldwide (including in Australia, the United Kingdom, and the United States (to name a few)), have accelerated efforts to spur widespread aircraft electrification, citing the move as an option for a "cleaner, faster and more convenient air transport (11,12)."

However, widespread deployment of e-aircraft is challenged in large measure by battery technology. Although electric motors are more efficient at converting electricity into propulsive force (compared to combustion engines powered by fossil fuels), this efficiency is insufficient to offset the low gravimetric energy density of the batteries that power these motors (5,6,10,13,14). This insufficiency impedes an electrified aircraft's ability to transport passengers, particularly over long distances where the associated battery mass renders such flights impractical. Consequently, existing discourse emphasizes e-aircrafts' potential to service markets located in close spatial proximity to one another as doing so tempers the energy requirement (and by consequence, mass of battery) necessitated by the flight (6,8,10,15,16).

Alleviating battery mass requirements also motivates – in existing literature - the electrification of smaller aircraft which are - compared to their larger counterparts – lighter (17). Given the relationship between requisite force and mass, lighter aircraft require less energy to fly (which further tempers the energy and consequently, battery mass requirements). Physics notwithstanding, electrifying short haul routes and serving these routes using smaller, lighter aircraft – typically turboprops and regional jets - have some operational justification. Shorter flights are often characterized by thin passenger flows that demand less seat capacity (on a per flight basis) (6). Fewer seats in turn, facilitates load factor maximization, which is crucial to attaining and maintaining profitability (18,19). But could smaller aircraft also embody performance characteristics, that - from the vantage point of electrification – impede (rather than facilitate) short haul electrification?

The mass of an aircraft powered by liquified fuel changes during flight as continuous fuel combustion makes the aircraft progressively lighter. Consequently, an aircraft weighs less during landing than it does during takeoff. Engineers account for this when establishing the maximum landing weight (MLW), the heaviest weight at which an aircraft is certified to land. A lower MLW reflects a deliberate attempt to ensure safe deceleration and protects the landing gear and brakes from landing-related stress and impact forces. Exceeding the MLW risks damaging the aircraft's structure and overrunning the runway as the braking performance of an aircraft is inhibited when the aircraft is heavier.

Exceeding the MLW is what electrification risks because a battery's state of charge does not affect its weight (20). Unlike liquified fuel powered aircraft, an aircraft powered by batteries weighs as much during landing as it does during takeoff. This is particularly problematic for smaller, lighter aircraft because their ability to carry batteries is more restricted by virtue of having a lower MLW. Hence, holding flight length constant, though smaller aircraft may require a smaller battery to execute a flight (compared to a larger aircraft), given that moving less mass requires less energy, the battery for a smaller aircraft mass potentially occupies a larger *proportion* of that aircraft's MLW. Consequently, from the vantage point of safely landing, electrification may be a larger impediment for smaller aircraft versus larger aircraft.

Does such reasoning have merit? If so, to what extent? Are there other aircraft models (i.e., certain categories of mainline aircraft) that – where MLW considerations are concerned – may be more appropriate for short haul flights compared to the aircraft models (turboprops or regional jets) emphasized in discourse today? If so, how many routes do these aircraft models currently serve and what is the associated emissions reduction potential? To date, no studies have – to our knowledge – addressed the questions. Yet doing so is timely given that the transportation sector is responsible for a quarter of global carbon emissions, and emissions from the aviation sector specifically have – in recent years - grown faster that other travel modes, like rail, road and shipping (21,22).

We do so here. Our study scrutinizes e-aviation's potential to reduce emissions in the short haul aviation market. We distinguish our efforts from those prior by assessing – rather than presuming – which aircraft categories and models may be appropriate for short haul route electrification given aircrafts' performance characteristics, specifically an aircraft's MLW. We note that – to our knowledge, - MLW remains an overlooked parameters in aircraft electrification discourse. Yet, it is crucial to ensuring that aircraft safety is preserved, regardless of propulsion type. Furthermore, unlike prior work, we enumerate the magnitude of emissions reductions that may accordingly ensue by considering not only which aircraft models may be most appropriate for electrification but also accounting for how often these aircraft are deployed on commercial routes and where these routes originate (23). Our efforts can inform ongoing decarbonization efforts that emphasize electrification as a pathway for the aviation sector.

**Method**

Our analysis consists of five steps. First, we enumerate – for different existing aircraft models spanning different existing aircraft categories (i.e., turboprop, regional jet, narrowbody and widebody) – the requisite battery weight for a short haul flight. Second, we scrutinize whether the accommodation of this weight—coupled with a full passenger complement—exceeds the aircraft's MLW. Third, we identify aircraft models that are most appropriate for electrification: i.e., instances where MLW exceedance is minimal as lower exceedance is indicative of a smaller requisite increase in energy density to facilitate electrification. Fourth, we assess – for aircraft whose MLW exceedance is minimal – global deployment frequency. This step is motivated by the idea that maximum emissions reductions are likely when electrification occurs on routes high frequency routes. Fifth and finally, we enumerate the emissions reductions associated with electrifying those routes.

We recognize that whether existing aircraft models will be electrified as opposed to new aircraft models being introduced remains unclear. Future electrified aircraft models could drastically diverge from fossil fuel powered aircraft models seen today. Nevertheless, we believe our approach – which scrutinizes categories and models of existing aircraft – is useful in understanding what characteristics of these aircraft may position them (or their derivatives) as being most (or least) appropriate for electrification. We further note that while aircraft design has changed (e.g., adoption of composite wing construction, fuselage strengthening, and engines with higher bypass ratios), these changes are largely evolutionary (rather than revolutionary). To the extent that some designs are revolutionary (e.g., strut-braced wings, boundary layer ingestion, and advanced turbofan to name a few)(24), rapid adoption is unlikely as airlines plan their growth projections years in advance and aircraft are ordered to meet future – not current – demand (25). This limits the near-term effectiveness of revolutionary technology, which is problematic when rapid decarbonization is the end goal (26).

*Step 1, Battery weight estimate*: To estimate battery weight, we first establish a 200 nautical miles threshold for a short haul route. This process is informed by scrutiny of high frequency service between closely spaced airports (e.g., Kuala Lumpur to Singapore (161 nautical miles), Sao Paulo to Rio (182 nautical miles), Jeju to Busan (225 nautical miles), and London to Paris (188 nautical miles) (to name a few)). Though we acknowledge that many high-frequency routes can exceed 200 nautical miles (27), we note that the 200 nautical miles is the current standard industry electrification goal: airlines have expressed their desire to operate flights of up to 200 nautical miles, and manufacturers have targeted this range in the production of e-aircraft (28,29)

For this flight length, we subsequently calculate the requisite fuel requirements using different aircraft models of fossil fuel kerosene powered aircraft for which fuel data is available (30-35). These models include turboprops (4 aircraft), regional jets (13 aircraft), narrowbodies (14 aircraft), and widebodies (16 aircraft). Fossil fuel kerosene requirements for each of these models are estimated and subsequently converted to battery mass by accounting for passenger load factor (100 percent), thermal efficiency of combustion and electric engines (40 and 80 percent respectively), and battery mass conversion factor of 300 watt-hours per kilo (Wh/kg) (36,37,38).

For example, assuming a 100 percent load factor, an Embraer 170 – a popular regional jet used to service short haul markets - requires 1,091kg of kerosene to fly 200 nautical miles. Assuming 43.1 megajoules per kilogram (36), this yields a requisite energy requirement of 47,022 megajoules to cover this distance. Given a thermal efficiency of 40 percent, 18,809 megajoules are used for propulsion purposes (the rest being lost through friction and/or heat) (26). Assuming no change in requisite megajoules required for electrified propulsion, and a thermal efficiency of 80 percent, 23,511 megajoules are required to fly a 200 nautical mile flight using an electrified powertrain. Assuming a conversion factor of 1 megajoule = 277 watt-hours and current energy density estimate of 300 Wh/kg, this yields a battery weight of 21,769 kg (see Supplementary Information (SI): Section A for detailed overview for all 47 aircraft models).

*Step 2, MLW exceedance assessment*: Having established the requisite battery weight for 47 difference aircraft models, we assess whether accommodation of this weight exceeds the MLW for each aircraft model. This is done by summing the requisite battery weight (estimated in Step 1), the empty operating weight of the aircraft (which includes seats and galley equipment but excludes fuel and passengers), and the weight of passengers and their cargo (assumed to be 95kg per passenger) (39). The result is subsequently divided by the MLW established by the aircraft manufacturer to yield an exceedance ratio. A ratio exceeding 1.0 indicate instances where electrification may compromise the structural integrity of the aircraft during landing. Conversely, ratios below 1.0 indicate electrification indicate – given our model assumptions – instances where electrification may be a plausible prospect.

*Step 3, Aircraft model identification*: Having established exceedance ratios for 47 different models of aircraft, we identify the aircraft that are most appropriate for electrification (i.e., those with the smallest exceedance ratios). We acknowledge a-priori the possibility that none of the 47 aircraft models may be appropriate for short haul route electrification. That is, the MLW of all 47 aircraft may – given existing energy density limitations - be exceeded owing to electrification. In this scenario, we 1) enumerate the requisite increase in energy density required for each aircraft model to complete the flight without exceeding the MLW, and 2) identify the aircraft models that require the smallest percentage increase in energy density to complete the flight without exceeding the MLW. This approach reflects the premise the smaller energy density increases may - given the current trajectory of battery technology - be easier to achieve than larger ones (40,41).

*Step 4, Aircraft route assessment*: Aircraft models with the least exceedance are subsequently further scrutinized to ascertain their deployment frequency (i.e., how often these aircraft are used annually to transport passengers). Deployment frequency is ascertained first, regardless of route distance, and second, only on routes consistent with our 200 nautical mile threshold. This approach reflects the capital-intensive nature of aircraft procurement and accommodates the premise that electrifying an entire fleet may be financial unviable and/or logistically impractical. Consequently, to the extent that electrification facilitates decarbonization, financial resources should be prioritized for aircraft models that, a) require the smallest increases in energy density *and*, b) yield maximum emissions reduction owing to high deployment frequency.

Deployment frequency is ascertained by leveraging annual commercial flight data from the Official Airline Guide (OAG), an air travel intelligence reference that aggregates data on airline schedules, cargo and aviation analytics (27). OAG's databases include flight information updated daily, worldwide flight schedules, origin/destination information, flight details, airline code, airport, and aircraft model.

We use 2019 as our target year, as it precedes the COVID-19 pandemic during which air travel demand collapsed and because since 2019, air travel demand has not fully recovered to 2019 levels. Consequently, 2019 provides – we argue – a more comprehensive and estimate of air travel demand unaffected by the pandemic. While the full OAG database contains 48,203,125 trips for 2019 (42), we exclude other modes of transportation that are also included in the database. This includes limos

(1,926), buses (485,770), trains (2,844,077), helicopters (594,663), road feeder service (5,255,037) and freighter flights (602,374). This reduces the dataset to 38,419,278 observations. We subsequently also exclude observations of aircraft models that are not commonly used for commercial flights (2,436,131), those we do not have fuel data for (453,600) or those with labels that do not denote a specific aircraft model (2,028,052), such as "A320 family" or "Boeing 777 all pax models". Our final subset with 47 aircraft models consists of 33,501,495 flights, representing 87 percent of all scheduled commercial passenger flights in 2019. Of this subset, 4,364,491 flights are below 200 nautical miles (see SI: Section B for distributional representation of aircraft deployment by flight distance).

*Step 5, Estimated emissions reductions*: Having identified specific aircraft models that are most appropriate for electrification (i.e., aircraft models that meet our criteria of minimal MLW exceedance and high deployment frequency), we subsequently estimate the potential emissions reductions associated with electrifying these models. We do so by, a) enumerating the number of 'electrification miles' (defined here as the aggregate miles in 2019 covered by these models for flights covering less than 200 nautical miles), and b) enumerating the emissions footprint associated with covering these miles using fossil fuel kerosene versus electric propulsion.

Fuel consumption estimates for specified aircraft models are derived by plotting fuel usage against route distance, generating trend lines that inform fuel requirements as a function of distance. These fuel requirements are then – accounting for the thermal efficiency of combustion and electric engines - converted into watt-hours, producing trend lines for battery-electric operations. Emissions for battery-electric operations are, on a per flight basis, subsequently estimated by leveraging watt-hour trend line data and country-specific grid carbon intensity values (kg $CO_2$e per watt-hour) based on the departure location for each route (43). Leveraging the carbon intensity of the departure location assumes that battery charging for an aircraft will occur at the aircraft's departure point.

**Results and Discussion**

Existing discourse emphasizes electrifying short haul routes using smaller, lighter aircraft as an important pathway towards decarbonizing the aviation sector. This emphasis reflects, 1) limitations in the energy density of batteries which informs serving markets located in close spatial proximity to one another, and 2) a need to temper requisite energy (and by consequence battery mass requirements) requirements, given the relationship between requisite force and aircraft mass. Turboprops and regional jets have long been identified as aircraft models that best meet this requirement given that they primarily service short haul flights and are lighter than their narrow and widebody counterparts. Leveraging performance data for 47 different aircraft models, we scrutinize the extent to which such reasoning has merit and what the impact on emissions reductions are. Unlike previous efforts, we consider MLW thresholds, recognizing that exceeding these thresholds risks compromising the structural integrity of the aircraft.

Our analysis yields three key findings.

First, we find that – given the current day energy density profile of batteries - electrification prospects for short haul travel are impeded, regardless of aircraft model. After accounting for an aircraft's empty operating weight, passengers and cargo weight, and the weight of the battery that exhibits an energy density of 300 Wh/kg, the MLW is exceeded for all 47 aircraft models in our model (Fig 1a). The requisite energy density required to remain with MLW tolerance ranges from 461 Wh/kg (a 53.7 percent increase) for the Boeing 789-9, to 3,089 Wh/kg (a 1,039 percent increase) for the Dornier 328, with the average being 1,400 Wh/kg (a 467 percent increase). We note that these estimates exceed preceding enumerations of requisite energy density seen as being achievable over the next decade given sufficient investment in aeronautical applications (44). This excess reflects historical emphasis placed on adhering to the maximum takeoff weight (MTOW) (45) which is higher than the maximum landing weight and consequently imposes a less stringent energy density burden. Aggregated across all aircraft types, conformance to the MTO requires an energy density of 693.91 Wh/kg versus 1400 Wh/kg for the MLW. Nevertheless, our figures, which exceed those seen in practical lithium-ion batteries today (46), are consistent with longstanding literature that, 1) identifies energy density as an important impediment to aircraft electrification (5,10), and 2) emphasizes the success of aircraft electrification as being dependent – in part - on improvements in energy density.

However, unlike previous work, we find that turboprops and regional jets may be less appropriate for electrification compared to their narrow and widebody counterparts. Turboprops and regional jets exhibit the highest exceedance, the average being 1.72 and 1.66 respectively, compared to narrowbodies and widebodies which demonstrate an average exceedance of 1.47 and 1.30 respectively. Given the relationship between exceedance and requisite energy density (higher exceedance necessitates high energy density to remain within the MLW specified by manufacturer), completion of a 200 nautical mile flight necessitates that batteries for turboprops and regional jets exhibit higher energy density (2,144 Wh/kg and 1,979 Wh/kg respectively), compared to narrowbodies and widebodies for which the requisite energy density is 1,314 Wh/kg and 818 Wh/kg respectively. This finding supports our supposition that because the battery mass for a smaller aircraft potentially occupies a larger *proportion* of that aircraft's MLW, electrification may pose a greater risk for smaller versus larger aircraft.

While scrutiny of MLW exceedance helps identify aircraft models that may – given short haul travel - be more (versus less) appropriate for electrification, exceedance alone cannot be the sole determinant of electrification. The capital-intensive nature of aircraft procurement and operation makes electrifying every aircraft with favorable exceedance financially unviable and/or logistically impractical. This sentiment is reflected in existing commercial aviation operations as airlines routinely fly a combination of modern, more fuel-efficient aircraft models alongside older, less efficient ones (47). Consequently, of relevance is not only which aircraft demonstrate the lowest exceedance, but also how frequently these aircraft are deployed on short haul routes. We scrutinize deployment frequency by analyzing 33,501,495 commercial scheduled flights in 2019, 4,364,491 of which meet our 200 nautical mile threshold.

Considering deployment frequency concurrently with MLW exceedance elucidates our second finding. We find that whereas widebody aircraft demonstrate – on average – the lowest MLW exceedance (1.32), specific narrowbody aircraft may - given concurrent consideration of deployment frequency and MLW exceedance – be more appropriate for electrification (Fig. 1b). We find that of the 47 aircraft in our model, three narrowbody aircraft models, namely the Airbus A319, Airbus A320, and Airbus A321, demonstrate low exceedance (an average of 1.32) and high deployment frequency, collectively accounting for 885,894 of 4,364,491 flights (20.3 percent) in our short haul flight sample. The Airbus A319 has an exceedance of 1.30 and accounts for 202,777 flights (4.65 percent of flights under the 200 nautical mile threshold), the Airbus A320 has an exceedance of 1.30 and accounts for 547,247 flights (12.54 percent), and the Airbus 321 has an exceedance of 1.37 and accounts for 135,870 flights (3.11 percent). By comparison, the 44 other aircraft models demonstrate an average exceedance of 1.50 and account for 79.7 percent of flights that cover less than 200 nautical miles (3,478,597 of 4,364,491 flights).

Our finding is noteworthy given historical emphasis on turboprops and regional jets as being the aircraft models most appropriate for electrifying short haul routes. We note that while this aircraft model choice (i.e., smaller, lighter aircraft) tempers the energy requirement and by consequence, mass of battery necessitated to complete the flight (compared to larger, heavier aircraft that require more energy and consequently, larger batteries), smaller, lighter aircraft have more restrictive MLW requirements (compared to larger, heavier aircraft) which may make them – given current limitations in energy density – less appropriate for electrification. Our results suggest that larger aircraft, specifically some models of narrowbody aircraft, serving short haul routes may – given their flight performance profile – be more appropriate for electrification compared to their turboprop and regional counterparts.

We recognize that this approach implies emission reductions for a minority of aircraft models that collectively account for a minority of annual short haul flights. We identify 44 aircraft models (out of 47) as being less appropriate – from the vantage point of MLW exceedance – for electrification and these models account for 3,478,597 flights annually (out of 4,364,491 flights). Given the timeliness of tempering emissions in the aviation sector, some may argue for electrification of all aircraft models servicing short haul routes, rather than a subset of aircraft deployed on these routes. However, as previously noted, such reasoning ignores, a) the magnitude of energy density improvement that is required by the aircraft in our model to remain within MLW tolerances, b) the capital-intensive nature of the aviation sector that – assuming energy density were not an impediment – makes complete fleet electrification challenging, and c) the three aircraft models identified for electrification 'punch above their weight' in terms of deployment frequency (i.e., 6.4 percent of aircraft models in our model collectively account for 20.3 percent of flights covering less than 200 nautical miles).

What are the potential emissions benefits of deploying electrified A319,A320, and A321 aircraft on short haul routes? Our third finding is that at least 917,826,722 kg $CO_2$e may be avoided annually by electrifying these aircraft models[1]. This estimate is informed by considering the distance of all commercially scheduled flights flown by these aircraft models in 2019 (Fig. 2a) and scrutinizing specific flights covering less than 200 nautical miles (Fig. 2b). For these flights, the requisite energy requirement (and subsequent emissions product) is estimated given this distance threshold, aircraft model, and the departure point of the aircraft (which informs the carbon intensity of the grid for battery-electric aircraft) (43). At a regional level, the largest emissions benefit is realized by electrifying flights originating in Europe (533,101,759 kg $CO_2$e avoided), followed by South America (433,529,588 kg $CO_2$e avoided), North America (104,850,070 kg $CO_2$e avoided), Oceania (12,582,047 kg $CO_2$e avoided), and Africa (2,386,736 kg $CO_2$e avoided) (Fig. 3a). We note that electrifying flights originating in Asia produces an increase in carbon emissions owing to electrification (168,623,478 kg $CO_2$e produced).

How might these results be explained? What makes Europe a better candidate for aircraft electrification compared to other regions? And why does electrifying routes originating in Asia produce an increase in emissions? Our regional emissions benefit breakdown lacks context absent consideration of, a) the number of flights originating in each of these regions, and b) the carbon intensity of countries within each of these regions. For example, large emissions benefits seen in Europe may reflect the deployment of more A319/320/321s which subsequently offers great emissions savings compared to Asia which may have fewer of such aircraft operating on routes less than 200 nautical miles). Alternatively, larger emissions savings may also reflect availability of a less carbon intensive electrical grid compared to other regions. We scrutinize the legitimacy of these explanations by accounting for the total flight miles travelled in each of these regions by the specified aircraft model coupled with the regional carbon intensity of the electrical grid.

We find that Europe's position as demonstrating the greatest emissions benefit (533,101,759 kg CO2e) reflects both, a large number of electrification miles (36,702,126 nautical miles) and a less carbon intensive electrical grid (the continent-wide average being 311.72 g$CO_2$e/kWh). Conversely, the increase in carbon emissions observed in Asia reflects many electrification miles (61,470,196 nautical miles) and a far more carbon intensive grid (the continent-wide average being 553.63 g$CO_2$e/kWh). This produces an increase – rather than decrease – in Asia's emissions owing to electrification. Excluding Oceania, South America benefits from having the cleanest electrical grid (216.80 g$CO_2$e/kWh) but compared to Europe, South America's aggregate emissions reduction potential is lower (433,529,587 kg $CO_2$e avoided compared to 533,101,759 kg $CO_2$e avoided) owing to fewer electrification miles (23,886,445 nautical miles compared to Europe's 36,702,126 nautical miles).

Asia's emergence as a poor candidate for electrification warrants further scrutiny given this region is expected to account for over half of the world's passenger growth by 2043 (48). For the specific aircraft models and route type, Asia accounts for 44.82 percent of electrification miles available globally but is responsible for 98.42 percent of global carbon emissions produced owing to electrification (Fig. 3b). To understand why, we scrutinize the relationship between emissions savings and carbon intensity of the electrical grid at a country level. We find that countries with a grid carbon intensity higher than approximately 530g$CO_2$e/kWh (hereafter referred to as the 'tipping point') produce an increase in carbon emissions[2]. This is problematic for Asia because far more Asian countries have grid intensities that exceed this figure. The carbon intensity of 22 of 36 Asian countries (61.1 percent) exceeds 530 $gCO_2e/kWh$. By contrast, 7 of 13 countries in Africa (53.8 percent), 7 of 36 countries in Europe (19.4 percent), 4 of 12 countries in North America (33.3 percent), and 1 of 2 countries in Oceana (50 percent) have grid carbon intensities exceeding 530 $gCO_2e/kWh$ (see SI: Section C for country specific breakdown).

[1] This figure slightly understates the true benefit magnitude of electrification as it accounts for 5,937 fewer flights (879,530 versus 885,894) than those identified owing in large measure to the absence of reliable carbon intensity data for the electrical grids for specific regions (e.g., Macau and Jersey).

Asia's emissions increase also reflects - in large measure - the disproportionate share of electrification miles that are claimed by more polluting countries across the continent. Asian countries with a grid intensity exceeding 530 $gCO_2e/kWh$ account for 67.27 percent of continent-wide electrification miles (41,347,945 of 61,470,196 nautical miles) (Table 1). By contrast, European countries with a grid intensity exceeding 530 $gCO_2e/kWh$ account for 1.22 percent of continent-wide electrification miles (447,053 of 36,702,126 nautical miles) and North American countries with a grid intensity exceeding 530 $gCO_2e/kWh$ account for 0.80 percent of continent-wide electrification miles (109,796 of 13,731,056 nautical miles). Africa is an exception to this phenomenon as African countries with a grid intensity exceeding 530 $gCO_2e/kWh$ account for 68.92 percent of continent-wide electrification miles (514,866 of 747,102 nautical miles). However, the continent still generates an emissions decrease owing to electrification (2,386,736 kg $CO_2e$ avoided) as the miles flown for flights originating in cleaner African countries (i.e., those with a grid intensity lower than 530 $gCO_2e/kWh$) offer greater emissions reductions on a per mile basis (20.23 kg $CO_2e$) compared to per mile emissions generated by flights originating in dirtier African countries (4.49 kg $CO_22e$). This effect is the product of cleaner African countries having an average grid intensity that is 50.33 percent lower (263.25 $gCO_2e/kWh$) than the 530 $gCO_2e/kWh$ tipping point, compared to dirtier African countries whose average grid intensity is 18.82 percent higher (629.74 $gCO_2e/kWh$) than the 530 $gCO_2/kWh$ tipping point.

At the country level, two India and Brazil – warrant discussion (Fig. 3b). Both counties are expected to see significant increase in air travel demand over the coming years but our analysis highlights potentially opposing emissions trajectories owing to electrification (49,50).

India emerges as being a prominent emitter, accounting for 15.45 percent of electrification miles across Asia (9,495,447 of 61,470,196 nautical miles) but responsible for 36.21 percent of the emissions (91,652,871 of 253,113,374 kg $CO_2e$) generated across the continent. These figures are even more pronounced at the global level. For our specified flight distance and aircraft models, India accounts for 6.92 percent of electrification miles globally (9,495,447 of 137,155,685 miles) but is responsible for 35.64 percent of the emissions produced (91,652,871 of 257,169,107 kg $CO_2e$). Contrastingly, Brazil emerges as the largest benefactor of electrification, delivering the largest emissions savings. Brazil accounts for 38.55 percent of electrification miles across South America (9,208,587 of 23,886,445 nautical miles) but is responsible for 45.86 percent of the emissions reduced (198,824,978 of 433,529,588 kg $CO_2e$ avoided) across the continent. These figures are also pronounced at the global level (albeit not to the same extent as India). Brazil accounts for just 6.71 percent of electrification miles globally (9,208,587 of 137,155,685 miles) but is responsible for 21.66 percent of the emissions reduced (198,824,978 of 917,826,722 kg $CO_2$e).

[2] We determine this tipping point by identifying the grid carbon intensity that would equalize emissions generated by fuel and electricity-powered flights in a country. The tipping point is estimated for each country and ranges from 526.89 gCO2/kWh to 529.94 gCO2/kWh (see SI: Section D). Given this range, when engaging in cross continent comparison, we use 530 gCO2/kWh as the most conservative estimate of how `dirty' a grid must for emissions generated by fuel and electricity-powered flights to equalize.

What explains the diverging emissions trajectories of India and Brazil? Answering this question is facilitated by both countries offering – for our specified flight length and aircraft models – almost identical number of electrification miles (9,495,447 and 9,208,587 for India and Brazil respectively). Given comparable electrification miles, we find that India's position as an emissions producer is largely explained by a carbon grid intensity that exceeds the tipping point (compared to Brazil whose carbon grid intensity is below the tipping point, which results in an emissions reduction). However, we also observe that Brazil's electrical grid is far cleaner than India's is dirtier relative to the tipping point. Brazil has a carbon grid intensity that is 81.51 percent lower (98 g$CO_2$e/kWh) than the tipping point compared to India's carbon grid intensity which is 35 percent higher than the tipping point (713 g$CO_2$e/kWh). This relative difference explains in part why – despite comparable electrification miles – carbon emissions avoided is significantly higher in Brazil (198,824,978 kg $CO_2$e avoided) that carbon emissions produced in India (91,652,871 kg $CO_2$e produced)[3].

Our emphasis on India and Brazil should not detract from the need to also scrutinize the emissions contributions of other countries owing to electrification. Saudi Arabia, Malaysia, and Indonesia – the top three emissions producers after India - collectively account for 39.69 percent of global emissions produced. Conversely, Columbia, the United Kingdom, and Spain - the largest beneficiaries of electrification after Brazil - collectively account for 30.37 percent of global emissions avoided. Which countries benefit (and which ones do not) reflects first and foremost, heterogeneity in carbon grid intensity, but also – depending on the country – the number of electrification miles, the number of flights flown, and the model of aircraft flown. Consideration of these parameters by policymakers is timely as effects to decarbonize the aviation sector accelerate.

Finally, we note that a less carbon intensive grid is, from an emissions reduction perspective, advantageous, it also yields – should efforts to further temper the grid carbon intensity - declining marginal gains. Improving Brazil's electric grid delivers – in absolute and relative terms – fewer emissions reductions compared to improving India's electrical grid (see SI: Section E for. Detailed country level rankings of emissions savings). Marginal gains also explain why – despite both countries being carbon emitters in our model - improving Columbia versus Ecuador's electrical grid may be more advantageous. Consequently, to the extent that carbon emissions are ultimately universal, and financial resources to temper these emissions are limited (51), directing these resources towards grid decarbonization efforts in countries that have the most carbon intensive grids may be timely (versus in countries that either have less carbon intensive grids (but are still emitters), or countries that generate negative emissions (e.g., Brazil). We emphasize that decarbonizing India's power sector is – from the vantage point of aircraft electrification - particularly timely given India's expected population growth over the next decade. This growth is expected to be coupled with a 6.2 percent annual increase in passenger demand by 2040, well above the global average of 3.9 percent and the fastest among major economies (52). To the extent that a portion of this growth occurs in a 200 nautical mile market poised to electrify, emissions in this market could – absent efforts to decarbonize the power sector - rise. Although flights under 200 nautical miles that originate in India account for 8.69 percent of all flights (98,645 of 1,135,005), 64.07 percent of flights flying this flight profile are flown by our specified aircraft models (63,199 of 98,645 flights).

---

[3] Relative to Brazil, higher carbon emissions in India are also explained (albeit to a lesser extent than grid intensity) by the higher number of flights (63,199 versus 49,640) as – holding aggregate miles travelled constant across countries – more flights in a country will generate higher emissions than fewer flights in another country (see SI: Section F for details).

## Limitations and Conclusion

To assess the emissions reduction potential of electrified short haul flight, we scrutinize which aircraft models may be most appropriate for electrification. We subsequently consult publicly available route information to assess – for specified aircraft models - which regions may benefit from electrification and to what extent. Doing so necessitates leveraging of aircraft performance data, deployment patterns, and carbon grid intensity at the country level. Our model accounts for each of these parameters delivering – we argue – robust results. Nevertheless, limitations of our work warrant acknowledgment.

First, while the subset of narrowbody aircraft identified in our analysis - namely the Airbus A319, Airbus A320, and Airbus A321 reflect concurrent consideration of deployment frequency and MLW exceedance, we acknowledge that presence of other aircraft models in our model that also offer advantageous deployment frequency and more tolerable MLW exceedance. Specifically, the De Havilland Canada Dash 8 (a turboprop), Embraer 145 (a regional jet), and two additional narrowbodies, the Boeing 717-200, and the Boeing 737-800, also demonstrate low exceedance (an average of 1.48) and collectively account for 1,274,605 of 4,364,491 flights (29.2 percent) in our short haul flight sample (Fig 1a). The omission of these aircraft in our geographical analysis largely reflects the fact that, 1) the average exceedance is higher than that of the Airbus A319, Airbus A320, and Airbus A321 (which demonstrate and an average of exceedance of 1.32 while accounting for 20.3 percent of flights in our short haul flight sample), and 2) smaller energy density improvements may be easier to achieve than larger ones (40). We acknowledge that accounting for the four additional aircraft models specified above would increase the emissions reduction potential associated with electrifying short haul flights. We note however that inclusion of these four additional aircraft models does not change the regional effects observed (e.g., Asia being an emissions producer).

Secondly, although the identification of aircraft suitable for electrification assumes improvements in energy density will occur, we acknowledge that the precise trajectory of these improvements remains unclear. This lack of clarity influences – as previously noted and from the vantage point of exceedance – aircraft selection. Abrupt and significant improvements in energy density would admittedly make more aircraft models – particularly the turboprops and regional jets – more viable for electrification. However, we note that from a historical viewpoint, the energy density has never increased suddenly due to complicated system design and requirements on well-balanced performances for application and the average increasing rate of energy density of Li-ion batteries which has averaged three percent annually over the last quarter decade is decreasing (46,53). Furthermore, we observe that although many high-capacity batteries are being widely explored and advances have been made, the achieved energy density has generally not exceeded 300 Wh/kg and mainstream battery technology appears unable – for now – to continuously power large systems like aircraft (54,55). Consequently, our emphasis on aircraft requiring smaller energy density improvements is – we argue – timely.

Thirdly, our analysis assumes that electrification will not alleviate the weight imposed by other aircraft components. We acknowledge that were that not the case (i.e., weight reductions achieved by redesigning other aircraft components could offset weight demands of an electric battery), the aircraft's gross weight may be lower which would impose a less strict energy density requirement. However, evidence supporting such reasoning is lacking. Rather, existing evidence from other sectors such as the auto sector highlights the weight burden imposed by electrification and – barring energy density improvements – the inability of producers to meaningfully alleviate this burden (56). No evidence – to our knowledge – suggests a differing outcome for aviation. Moreover, we note that existing literature emphasizes battery weight as being the primary impediment to electrified flight owing to limits in energy density. Such reasoning, and associated data, justifies – we argue – the assumptions leveraged by our model and results produced.

Fourth, the results we present exclude fuel reserves requirements associated with commercial air travel. Reserve Fuel represents the additional fuel carried by aircraft beyond the planned requirements for a flight, and it serves ,as a critical safety buffer for unforeseen circumstances (e.g., delays, diversions, and/or unexpected changes in flight conditions (57,58). Regulators typical require that reserve fuel accommodate an additional 30 to 45 minutes of flight at 'normal cruising speed.' We note that further stress testing of our model by considering energy increases of 33 percent, 66 percent, and 100 percent (all of which reflect consideration of fuel reserve requirements) do not change our results highlighting their robustness.

Fifth, and finally,  while our analysis is predicated on energy density increases helping aircraft remain within MLW tolerances, we recognize there may be other ways to achieve the same outcome. The most notable is increasing an aircraft's MLW  (independent of an energy density increase), which would reduce exceedance. Indeed, a sensitivity analysis (see SI: Section G) of our model parameters highlights the utility of increasing MLW tolerances compared to other pathways (e.g., increasing energy density, reducing passenger weight). While accommodating higher MLW warrants scrutiny, we caution that increasing MLW would - regardless of aircraft model – require new design approaches and/or significant structural strengthening for landing gear to withstand increased impact forces. Doing so risks imposing larger, heavier, and more complex designs that could ultimately impose a weight penalty (necessitating an even larger MLW increase). Nevertheless, future research should scrutinize this pathway's viability.

Aviation remains among the most energy-intense forms of consumption and has in the past been characterized by strong growth, with estimates that emissions have increased significantly 1960 and 2018 (4,59-61). Electrification offers a means of tempering this trajectory, offering by relative to fossil-fueled reference aircraft a reduction in $CO_2e$ emissions of up to 88 percent (10). Given limits in energy density, electrification has historically been favored to accommodate short haul flights flown by smaller, lighter aircraft (typically turboprops and regional jets). Our results challenge such reasoning. We demonstrate that while energy density limitations will need to be overcome regardless of aircraft model, some narrowbody aircraft may – given the smaller requisite energy density increase and higher deployment frequency - be more appropriate for doing so. Electrifying these aircraft models for short haul flights would offer significant global emissions reductions, the most pronounced reductions being in Europe, and the least pronounced across Asia. Collectively, our findings warrant scrutiny by policymakers given aviation's concurrent role as a facilitator of economy growth and emissions contributor (2,26).

**Acknowledgements**

The authors thank Nick Johnson and Steve Walsh for their assistance in preparing this manuscript.

**Data availability**

Data used for the study will be made available upon request.

**Funding**

The authors declare no funding sources.

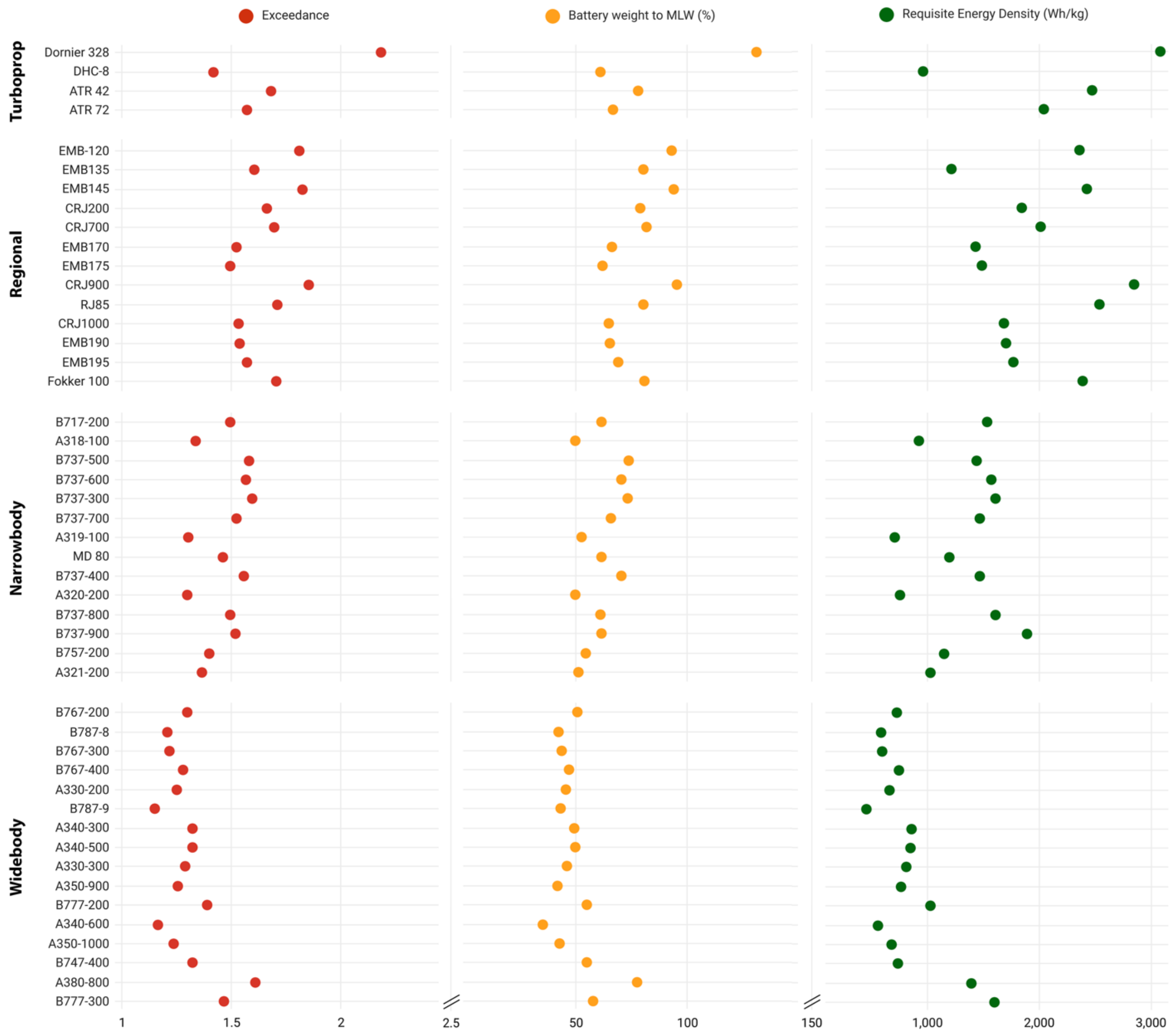

Figure 1a. Aircraft model exceedance ratio, battery to maximum landing weight ratio (%), and requisite energy density (Wh/kg)

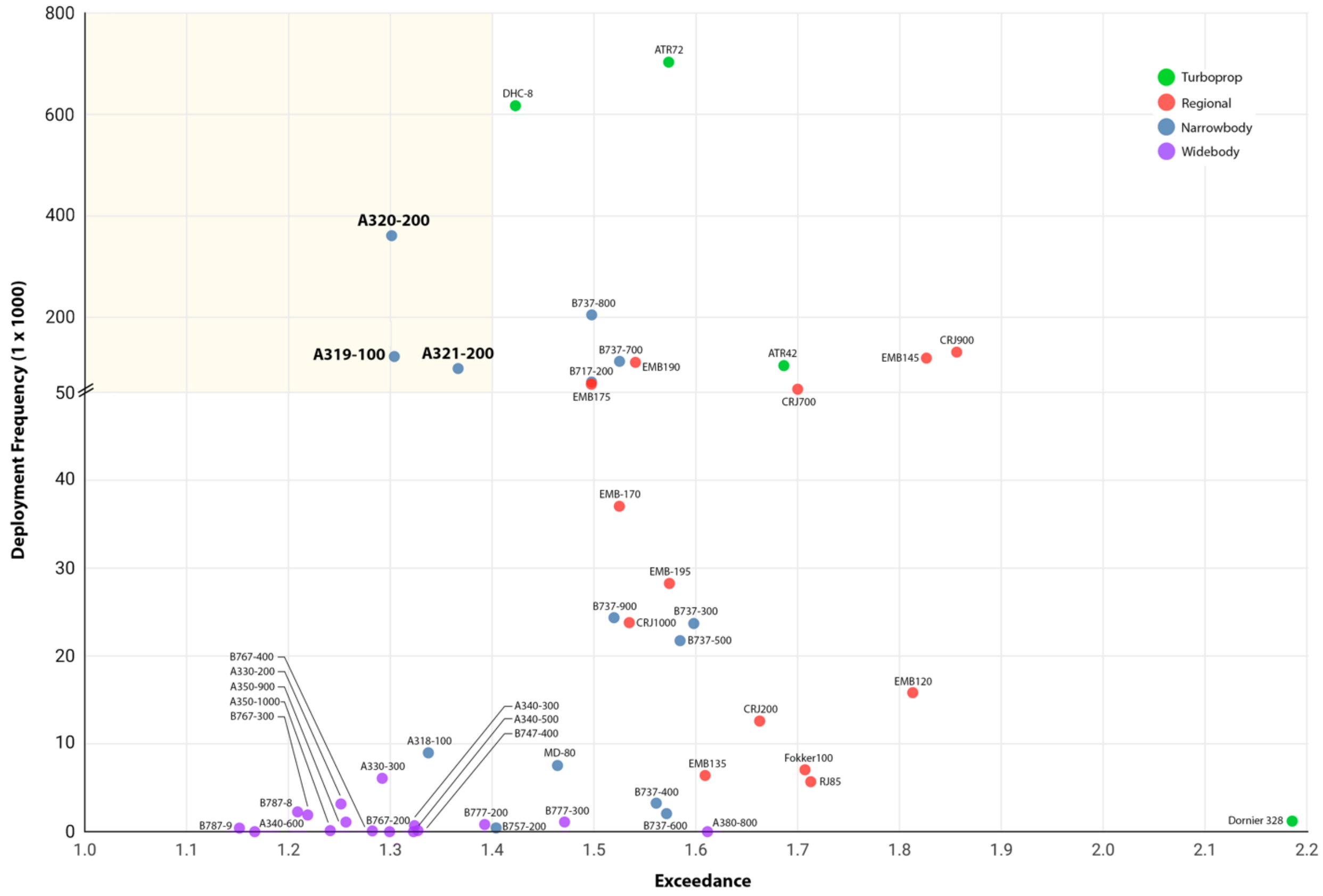

Figure 1b. Aircraft model exceedance ratio versus aircraft model deployment frequency

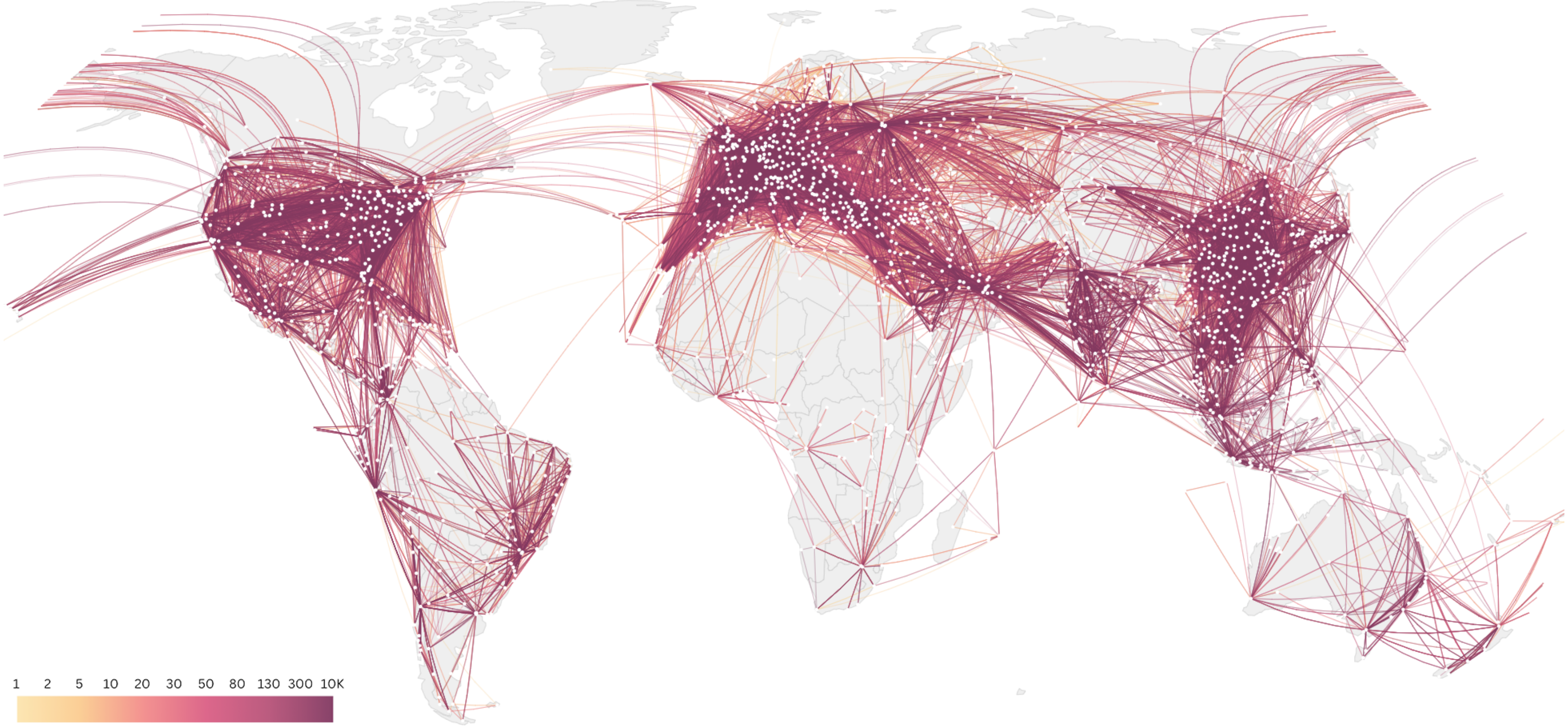

Fig 2a. Overview of 120,466 routes serviced by three aircraft models ((A319/A320/A321.) flying 12,726,937 flights in 2019. Deployment frequency is reflected by color intensity, which reflects decile increasements. For example, 10 percent of routes are served between 300 and 10,000 times annually (depicted by the darkest color). Conversely, 10 percent of routes are served between 1 and 2 times annually (depicted by the lightest color)

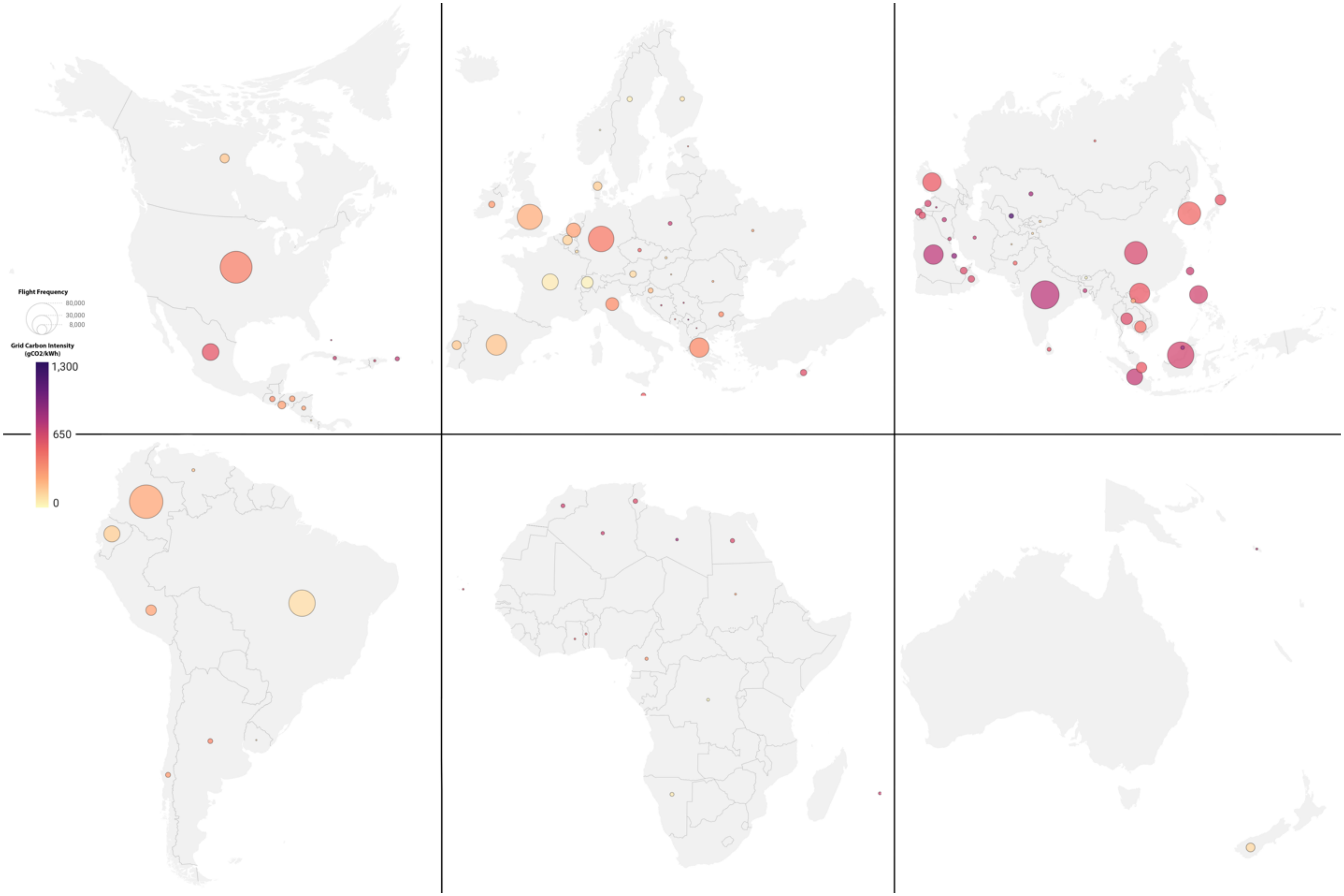

Fig 2b. Continent level overview of aircraft model deployment frequency and country level carbon grid intensity. For each continent, circle size reflects deployment frequency for all flights below the 200 nautical mile threshold for specified aircraft model (A319/A320/A321) and circle color reflects local carbon grid intensity.

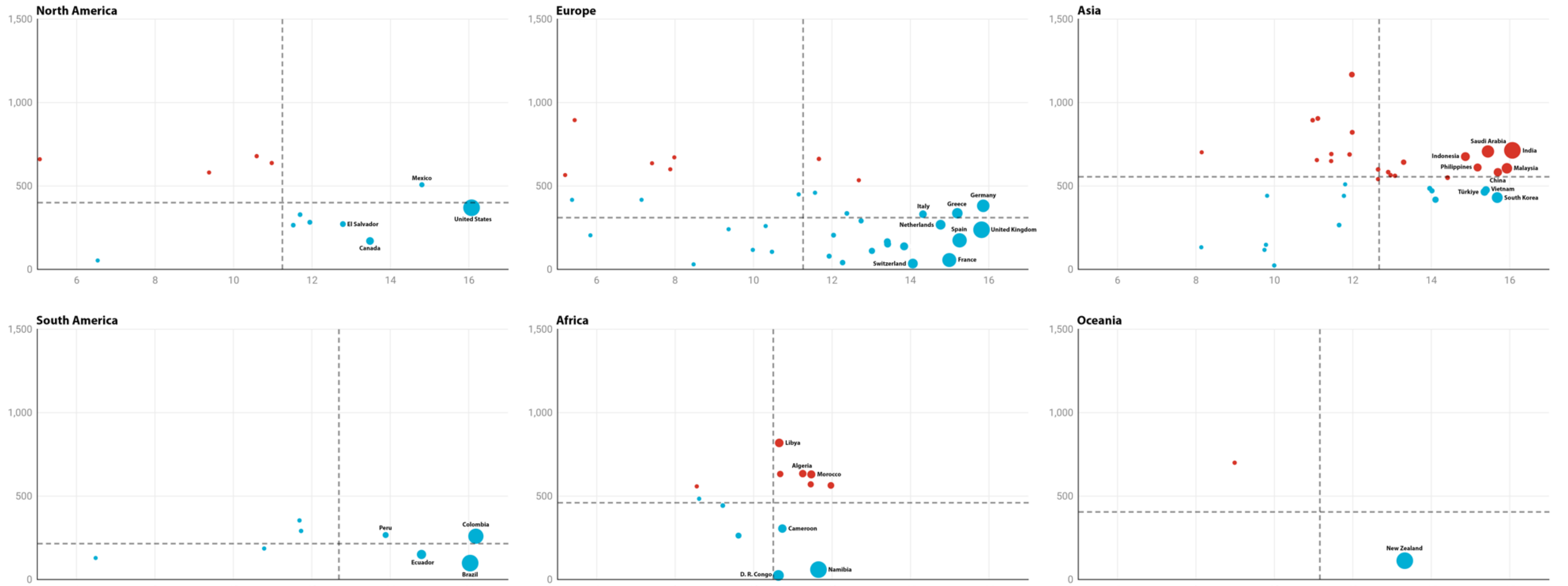

Fig 3a. Continent level emissions footprint owing to electrification. For each continent, blue circles are countries with an emissions reduction while red circles are countries with an emissions increase. Circle size reflects emissions magnitude (positive or negative) relative to other countries *within* the same continent. Vertical axis reflects grid carbon intensity (gCO2/kWh) and horizontal axis is the natural logarithm of available electrification miles for all flights below the 200 nautical mile threshold for specified aircraft models (A319/A320/A321). Axis within a continent reflect average grid carbon intensity and average available electrification miles across all countries within that continent.

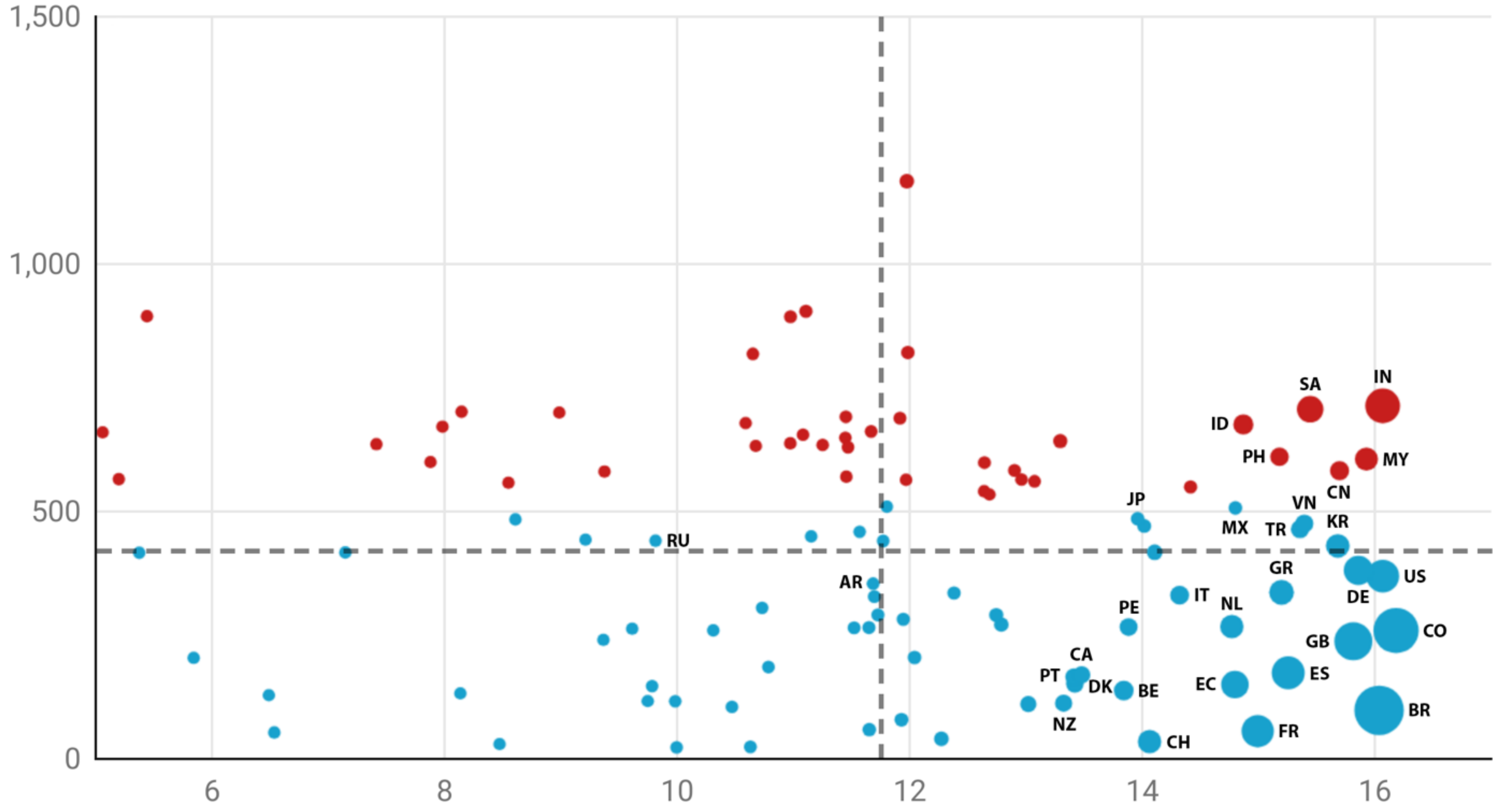

Fig 3b. Country level emissions footprint owing to electrification. Blue circles are countries with an emissions reduction while red circles are countries with an emissions increase. Circle size is indicative of emissions magnitude (positive or negative) relative to other countries. Vertical axis reflects grid carbon intensity (gCO2/kWh) and horizontal axis is the natural logarithm of available electrification miles for all flights below the 200 nautical mile threshold for specified aircraft models (A319/A320/A321). Dotted lines reflect average grid carbon intensity and average available electrification miles across all 105 countries. Labelled countries (31 of 105) are members of the G20 and/or countries whose emissions (positive or negative) exceed 10,000,000 kg CO2e.

| Continent | Grid Classification | Total Miles (nm) | Average Grid Intensity ($g\ CO_2e/kWh$) | Emissions Savings (kg CO2e) | Emissions Reduction per Mile Flown ($kg\ CO_2e/nm$) |
|---|---|---|---|---|---|
| Africa | Aggregate | 747,102 | 460.59 | 2,386,736 | 3.19 |
| | Clean | 232,236 | 263.25 | 4,699,689 | 20.24 |
| | Dirty | 514,866 | 629.74 | -2,312,952 | -4.49 |
| Asia | Aggregate | 61,470,196 | 553.63 | -168,623,478 | -2.74 |
| | Clean | 20,122,251 | 344.39 | 84,489,896 | 4.20 |
| | Dirty | 41,347,945 | 686.79 | -253,113,374 | -6.12 |
| Europe | Aggregate | 36,702,126 | 311.72 | 533,101,759 | 14.53 |
| | Clean | 36,255,073 | 223.51 | 534,038,660 | 14.73 |
| | Dirty | 447,053 | 651.96 | -936,901 | -2.10 |
| North America | Aggregate | 13,731,056 | 417.79 | 104,850,070 | 7.64 |
| | Clean | 13,621,260 | 280.91 | 105,588,360 | 7.75 |
| | Dirty | 109,796 | 639.31 | -738,289 | -6.72 |
| Oceania | Aggregate | 618,760 | 406.38 | 12,582,047 | 20.33 |
| | Clean | 610,777 | 112.76 | 12,649,638 | 20.71 |
| | Dirty | 7,982 | 700.00 | -67,591 | -8.47 |
| South America | Aggregate | 23,886,445 | 216.80 | 433,529,587 | 18.15 |
| | Clean | 23,886,445 | 216.80 | 433,529,587 | 18.15 |
| | Dirty | 0 | N/A | 0 | N/A |

Table 1. Continent breakdown of emissions profile owing to electrification. For each continent, emissions profile is further broken down by countries across the continent that are below ("clean") and above ("dirty") grid tipping point (530 gCO2/kWh).

**Supplementary Information Section**

In this section, we provide a summary of additional data/information that informs our model development.

*Section A: Aircraft model overview and operating parameters*

In Table S1, we summarize operating parameters and estimates of the fossil fuel and battery equivalent requirements to travel 200 nautical miles in one of 47 different aircraft models. Estimates assuming a 100 percent load factor, and a thermal efficiency for conventional and electric propulsion of 40 and 80 percent respectively. Battery weight estimates assume an energy density of 300 Wh/kg.

| Aircraft Model | Seating Capacity | Empty Weight (kg) | Total Passenger Weight (kg) | Total Fuel Weight (kg) | Total Battery Weight (kg) | Max. Takeoff Weight (kg) | Max. Landing Weight (kg) |
|---|---|---|---|---|---|---|---|
| EMB-120 | 30 | 7,070 | 2850 | 525 | 10,476 | 11,500 | 11,250 |
| Dornier 328 | 33 | 9,420 | 3135 | 947 | 18,896 | 15,660 | 14,390 |
| EMB135 | 37 | 11,308 | 3515 | 749 | 14,945 | 19,500 | 18,500 |
| DHC-8 | 39 | 9,979 | 3705 | 519 | 10,356 | 17,147 | 16,897 |
| ATR 42 | 48 | 10,285 | 4560 | 642 | 12,810 | 16,800 | 16,400 |
| EMB145 | 50 | 12,299 | 4750 | 912 | 18,198 | 23,050 | 19,300 |
| CRJ200 | 50 | 13,835 | 4750 | 845 | 16,861 | 24,041 | 21,319 |
| ATR 72 | 70 | 12,825 | 6650 | 727 | 14,506 | 21,750 | 21,600 |
| CRJ700 | 73 | 19,731 | 6935 | 1,252 | 24,982 | 32,995 | 30,390 |
| EMB170 | 80 | 20,646 | 7600 | 1,091 | 21,769 | 36,525 | 32,800 |
| EMB175 | 83 | 21,850 | 7885 | 1,061 | 21,171 | 39,580 | 34,000 |
| CRJ900 | 90 | 21,432 | 8550 | 1,599 | 31,906 | 36,514 | 33,345 |
| RJ85 | 99 | 23,882 | 9405 | 1,488 | 29,691 | 42,184 | 36,741 |
| CRJ1000 | 100 | 23,188 | 9500 | 1,205 | 24,044 | 41,232 | 36,968 |
| EMB190 | 106 | 27,959 | 10070 | 1,414 | 28,215 | 51,050 | 43,000 |
| B717-200 | 106 | 30,618 | 10070 | 1,433 | 28,594 | 53,524 | 46,266 |
| A318-100 | 112 | 37,000 | 10640 | 1,416 | 28,254 | 63,500 | 56,750 |
| B737-500 | 115 | 31,300 | 10925 | 1,846 | 36,835 | 60,555 | 49,895 |
| EMB195 | 115 | 28,819 | 10925 | 1,558 | 31,088 | 51,540 | 45,000 |
| Fokker 100 | 116 | 24,375 | 11020 | 1,595 | 31,826 | 43,770 | 39,348 |
| B737-600 | 119 | 36,378 | 11305 | 1,950 | 38,910 | 65,544 | 55,112 |
| B737-300 | 131 | 32,900 | 12445 | 1,934 | 38,590 | 63,277 | 52,527 |
| B737-700 | 138 | 37,648 | 13110 | 1,935 | 38,610 | 70,080 | 58,604 |
| A319-100 | 139 | 35,400 | 13205 | 1,647 | 32,864 | 75,500 | 62,450 |
| MD 80 | 152 | 35,380 | 14440 | 1,830 | 36,515 | 67,812 | 58,967 |
| B737-400 | 153 | 33,650 | 14535 | 1,988 | 39,668 | 68,039 | 56,246 |
| A320-200 | 165 | 37,320 | 15675 | 1,647 | 32,864 | 78,000 | 65,950 |
| B737-800 | 172 | 41,567 | 16340 | 2,000 | 39,907 | 74,388 | 65,314 |
| B737-900 | 183 | 42,493 | 17385 | 2,053 | 40,965 | 85,130 | 66,361 |

| Aircraft Model | Seating Capacity | Empty Weight (kg) | Total Passenger Weight (kg) | Total Fuel Weight (kg) | Total Battery Weight (kg) | Max. Takeoff Weight (kg) | Max. Landing Weight (kg) |
|---|---|---|---|---|---|---|---|
| B757-200 | 186 | 59,350 | 17670 | 2,458 | 49,046 | 104,350 | 89,800 |
| A321-200 | 203 | 47,500 | 19285 | 2,031 | 40,526 | 95,000 | 78,500 |
| B767-200 | 216 | 80,603 | 20520 | 3,270 | 65,249 | 136,083 | 128,054 |
| B787-8 | 242 | 112,050 | 22990 | 3,675 | 73,330 | 227,465 | 172,365 |
| B767-300 | 261 | 88,469 | 24795 | 3,190 | 63,652 | 181,437 | 145,150 |
| B767-400 | 270 | 103,147 | 25650 | 3,750 | 74,826 | 204,116 | 158,757 |
| A330-200 | 273 | 119,600 | 25935 | 4,184 | 83,486 | 242,000 | 183,000 |
| B787-9 | 290 | 110,677 | 27550 | 4,200 | 83,806 | 252,650 | 192,776 |
| A340-300 | 295 | 131,000 | 28025 | 4,767 | 95,119 | 275,750 | 192,000 |
| A340-500 | 313 | 168,000 | 29735 | 6,000 | 119,722 | 374,000 | 240,000 |
| A330-300 | 316 | 123,100 | 30020 | 4,275 | 85,302 | 233,000 | 184,500 |
| A350-900 | 325 | 142,400 | 30875 | 4,350 | 86,799 | 283,000 | 207,000 |
| B777-200 | 340 | 136,913 | 32300 | 5,604 | 111,821 | 236,096 | 201,800 |
| A340-600 | 380 | 174,000 | 36100 | 4,616 | 92,106 | 374,000 | 259,000 |
| A350-1000 | 380 | 155,000 | 36100 | 5,100 | 101,764 | 322,000 | 236,000 |
| B747-400 | 470 | 184,567 | 44650 | 8,181 | 163,241 | 412,770 | 295,743 |
| A380-800 | 525 | 276,791 | 49875 | 15,333 | 305,950 | 572,078 | 392,586 |
| B777-300 | 550 | 159,570 | 52250 | 6,907 | 137,820 | 299,370 | 237,680 |

Table S1: Aircraft model overview and operating parameters

*Section B: Aircraft model deployment frequency by flight distance*

Deployment frequency is ascertained by scrutinizing commercially scheduled flights using the OAG historical flight schedule database. For a given year, 2019, the database contains information for 48,203,125 trips. After excluding trips executed by limos (1,926), buses (485,770), trains (2,844,077), helicopters (594,663), road feeder service (5,255,037) freighter flights (602,374), aircraft models that are not commonly used for commercial flights (2,436,131), aircraft for which fuel data was unavailable (453,600) or those with labels that do not denote a specific aircraft model (2,028,052), such as “A320 family” or “Boeing 777 all pax models,” we arrive at a final data set of 33,501,495 trips executed by 47 aircraft models in 2019 (Fig. S1a).  Of this subset, 4,364,491 flights (13 percent) are below 200 nautical miles, and 885,894 of these 4,364,491 flights (20.29 percent) are assigned to the specified aircraft model type (A319,A320, A321) (Fig. S1b).

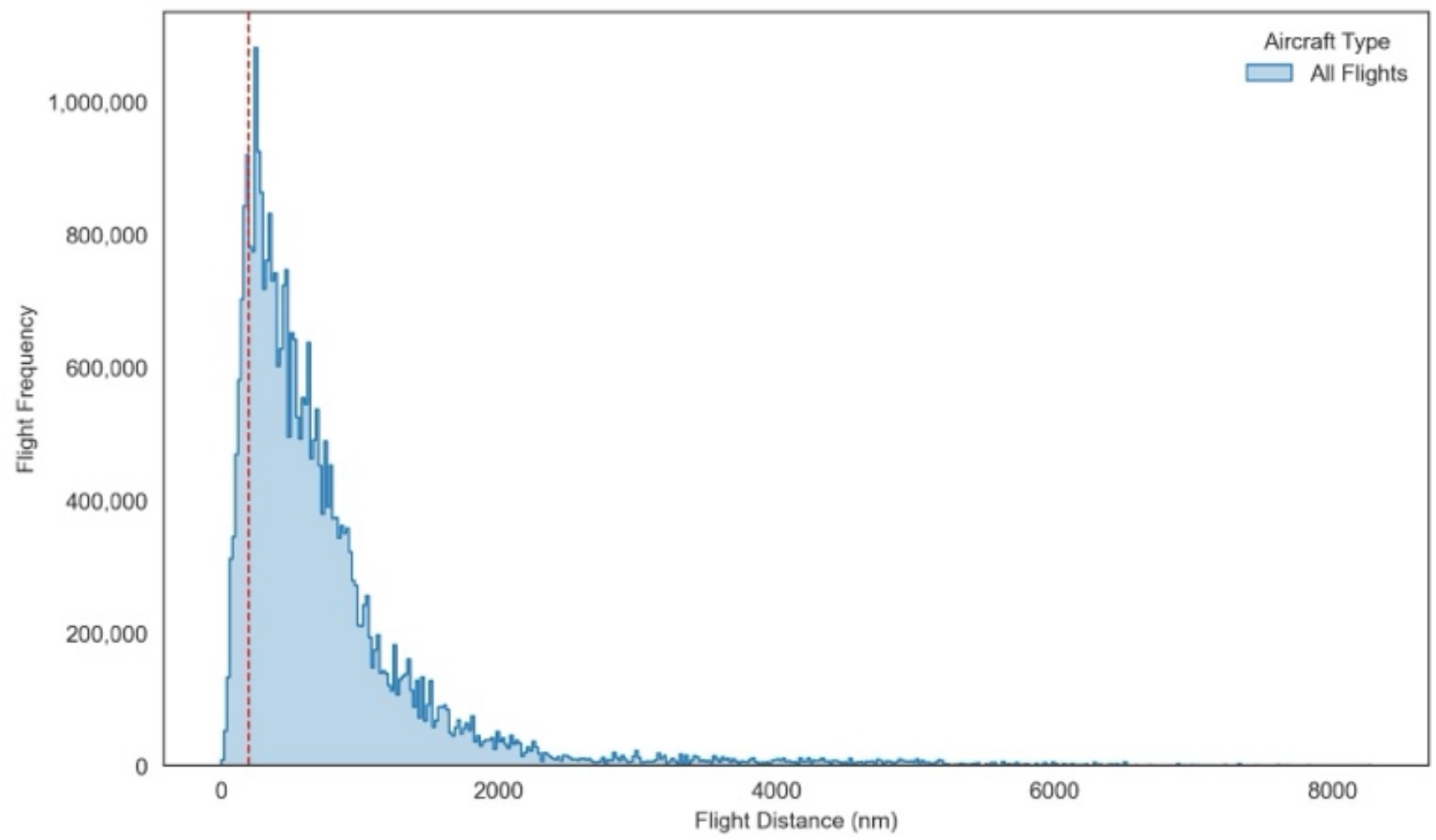

Figure  S1a: Distribution of aircraft model deployment frequency by flight distance

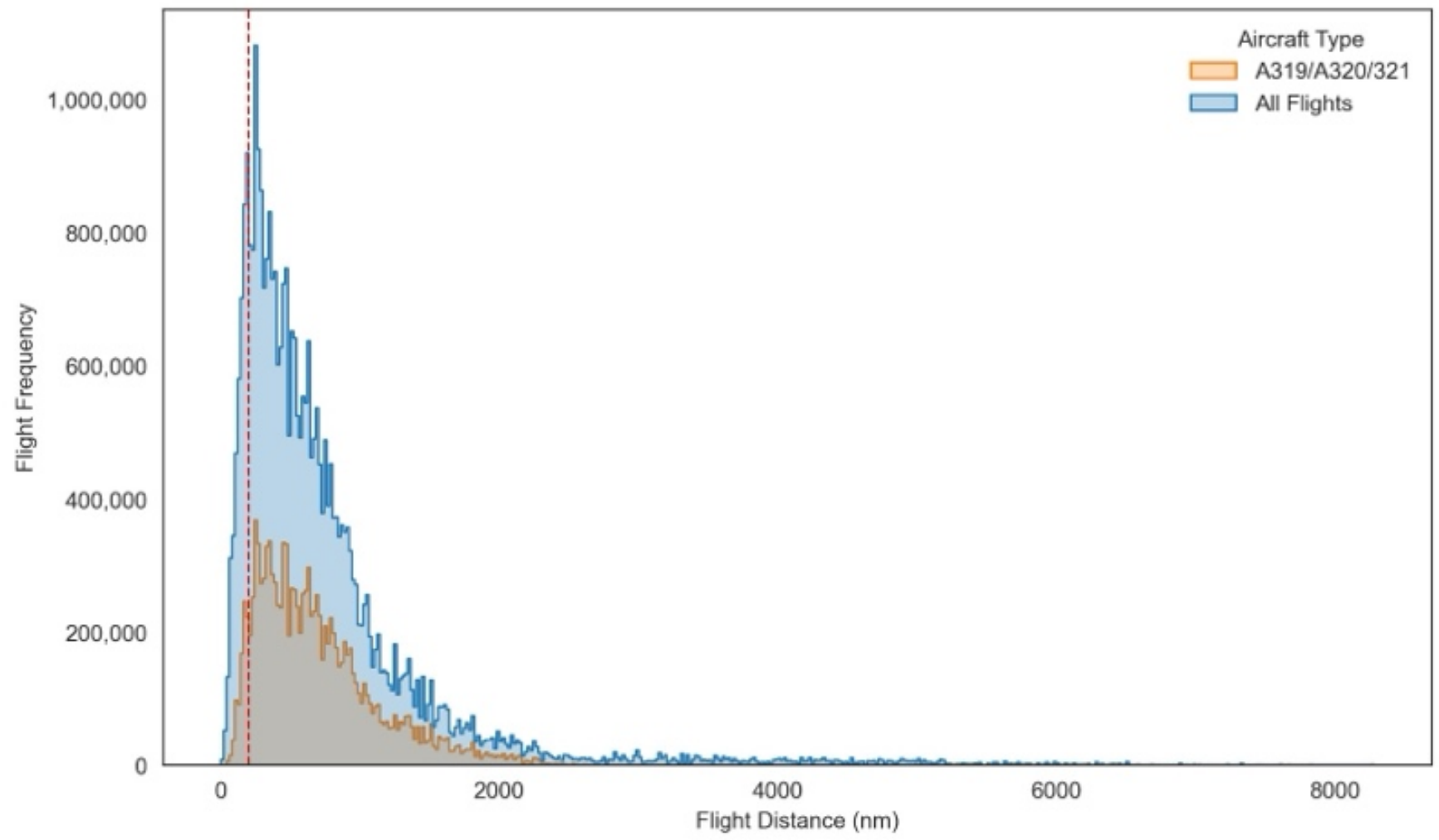

Figure  S1b: Distribution of aircraft model deployment frequency by flight distance with specified aircraft model breakdown

*Section C: Summary of country / continent level carbon intensity and available electrification miles*

SI Table 2 enumerates the carbon intensity of the 105 countries in our model, the absolute number of miles available for electrification (based on the specified aircraft models and distance threshold), and the associated emissions savings. Available electrification miles (and the associated emissions savings/costs) are expressed as a percent of both continent-wide miles and global miles. Countries are listed in order of emissions savings with those offering the most emissions savings owing to electrification appearing first (in this case, Brazil) and those offering the least emissions savings owing to electrification appearing last (in this case, India). Electrification of the specified aircraft models on routes less than 200 nautical miles causes an emissions decrease in 64 countries and an emissions increase in 41 countries.

| Country | Continent | Grid Intensity ($gCO_2e$ /kWh) | Tipping Point ($gCO_2e$ /kWh) | Available Electrification Miles (nm) | Continent Miles (%) | Global Miles (%) | Emissions Savings (kg $CO_2e$) | Continent Emissions Change (%) | Global Emissions Change (%) |
|---|---|---|---|---|---|---|---|---|---|
| Brazil | South America | 98.35 | 527.99 | 9,208,587 | 38.55 | 6.71 | 198,824,978 | 45.86 | 16.92 |
| Colombia | South America | 259.51 | 527.94 | 10,654,243 | 44.6 | 7.77 | 164,007,724 | 37.83 | 13.96 |
| United Kingdom | Europe | 237.59 | 528 | 7,377,915 | 20.1 | 5.38 | 111,902,200 | 20.95 | 9.52 |
| Spain | Europe | 174.05 | 527.99 | 4,214,480 | 11.48 | 3.07 | 80,886,880 | 15.15 | 6.88 |
| United States | North America | 369.47 | 527.99 | 9,499,961 | 69.19 | 6.93 | 79,943,985 | 75.71 | 6.8 |
| France | Europe | 56.04 | 527.69 | 3,241,619 | 8.83 | 2.36 | 77,224,991 | 14.46 | 6.57 |
| Germany | Europe | 380.95 | 527.93 | 7,702,483 | 20.99 | 5.62 | 60,529,793 | 11.33 | 5.15 |
| Ecuador | South America | 150.22 | 528.13 | 2,664,525 | 11.15 | 1.94 | 52,505,436 | 12.11 | 4.47 |
| Greece | Europe | 336.57 | 528.25 | 3,978,086 | 10.84 | 2.9 | 40,066,534 | 7.5 | 3.41 |
| South Korea | Asia | 430.57 | 528.75 | 6,451,546 | 10.5 | 4.7 | 35,519,800 | 42.04 | 3.02 |
| Switzerland | Europe | 34.84 | 528.33 | 1,279,215 | 3.49 | 0.93 | 34,478,796 | 6.46 | 2.93 |
| Netherlands | Europe | 267.62 | 528.42 | 2,594,662 | 7.07 | 1.89 | 34,114,344 | 6.39 | 2.9 |
| Belgium | Europe | 138.11 | 527.71 | 1,024,488 | 2.79 | 0.75 | 20,177,007 | 3.78 | 1.72 |
| Italy | Europe | 330.72 | 527.55 | 1,655,834 | 4.51 | 1.21 | 17,415,650 | 3.26 | 1.48 |
| Turkey | Asia | 464.59 | 528.24 | 4,663,745 | 7.59 | 3.4 | 15,559,096 | 18.42 | 1.32 |
| Vietnam | Asia | 475.45 | 527.49 | 4,839,339 | 7.87 | 3.53 | 15,193,303 | 17.98 | 1.29 |
| Peru | South America | 266.48 | 528.47 | 1,067,850 | 4.47 | 0.78 | 14,486,924 | 3.34 | 1.23 |
| Denmark | Europe | 151.65 | 528.05 | 672,724 | 1.83 | 0.49 | 13,564,173 | 2.54 | 1.15 |
| Portugal | Europe | 165.55 | 527.89 | 667,796 | 1.82 | 0.49 | 13,240,821 | 2.48 | 1.13 |
| Canada | North America | 170.04 | 527.13 | 713,001 | 5.19 | 0.52 | 13,022,219 | 12.33 | 1.11 |
| New Zealand | Oceania | 112.76 | 528.86 | 610,777 | 98.71 | 0.45 | 12,649,638 | 100 | 1.08 |
| Austria | Europe | 110.81 | 528.26 | 450,795 | 1.23 | 0.33 | 9,371,865 | 1.75 | 0.8 |
| Cambodia | Asia | 417.71 | 528.34 | 1,336,151 | 2.17 | 0.97 | 7,914,403 | 9.37 | 0.67 |
| Sweden | Europe | 40.69 | 527.59 | 213,439 | 0.58 | 0.16 | 5,433,512 | 1.02 | 0.46 |

| Country | Continent | Grid Intensity ($gCO_2e$ /kWh) | Tipping Point ($gCO_2e$ /kWh) | Available Electrification Miles (nm) | Continent Miles (%) | Global Miles (%) | Emissions Savings (kg $CO_2e$) | Continent Emissions Change (%) | Global Emissions Change (%) |
|---|---|---|---|---|---|---|---|---|---|
| El Salvador | North America | 271.47 | 528.91 | 357,345 | 2.6 | 0.26 | 5,002,510 | 4.74 | 0.43 |
| Ireland | Europe | 290.81 | 527.63 | 342,134 | 0.93 | 0.25 | 4,121,577 | 0.77 | 0.35 |
| Finland | Europe | 79.16 | 527.72 | 151,448 | 0.41 | 0.11 | 3,625,747 | 0.68 | 0.31 |
| Singapore | Asia | 470.78 | 528.39 | 1,221,516 | 1.99 | 0.89 | 3,557,615 | 4.21 | 0.3 |
| Croatia | Europe | 204.96 | 528.31 | 169,482 | 0.46 | 0.12 | 2,945,408 | 0.55 | 0.25 |
| Japan | Asia | 485.39 | 527.95 | 1,154,110 | 1.88 | 0.84 | 2,843,582 | 3.37 | 0.24 |
| Mexico | North America | 507.25 | 528.8 | 2,675,714 | 19.49 | 1.95 | 2,816,942 | 2.67 | 0.24 |
| Namibia | Africa | 59.26 | 527.74 | 114,814 | 15.37 | 0.08 | 2,792,410 | 59.42 | 0.24 |
| Bulgaria | Europe | 335.33 | 528.62 | 237,961 | 0.65 | 0.17 | 2,283,059 | 0.43 | 0.19 |
| Honduras | North America | 282.26 | 528.69 | 153,856 | 1.12 | 0.11 | 2,047,039 | 1.94 | 0.17 |
| Chile | South America | 291.11 | 528.67 | 123,660 | 0.52 | 0.09 | 1,681,657 | 0.39 | 0.14 |
| Laos | Asia | 265.51 | 528.86 | 114,482 | 0.19 | 0.08 | 1,640,094 | 1.94 | 0.14 |
| Guatemala | North America | 328.27 | 529.75 | 119,929 | 0.87 | 0.09 | 1,377,309 | 1.3 | 0.12 |
| Nicaragua | North America | 265.12 | 528.09 | 100,766 | 0.73 | 0.07 | 1,352,898 | 1.28 | 0.12 |
| Argentina | South America | 354.1 | 528.71 | 118,683 | 0.5 | 0.09 | 1,182,491 | 0.27 | 0.1 |
| D. R. Congo | Africa | 24.46 | 528.21 | 41,308 | 5.53 | 0.03 | 1,067,539 | 22.72 | 0.09 |
| Venezuela | South America | 185.8 | 527.65 | 48,240 | 0.2 | 0.04 | 827,215 | 0.19 | 0.07 |
| Luxembourg | Europe | 105.26 | 527.61 | 35,251 | 0.1 | 0.03 | 800,850 | 0.15 | 0.07 |
| Bhutan | Asia | 23.33 | 527.96 | 21,902 | 0.04 | 0.02 | 657,577 | 0.78 | 0.06 |
| Cameroon | Africa | 305.42 | 529.27 | 45,689 | 6.12 | 0.03 | 578,524 | 12.31 | 0.05 |
| Pakistan | Asia | 440.61 | 528.73 | 129,376 | 0.21 | 0.09 | 565,444 | 0.67 | 0.05 |
| Slovakia | Europe | 116.77 | 527 | 21,638 | 0.06 | 0.02 | 457,617 | 0.09 | 0.04 |
| Malta | Europe | 459.14 | 528.12 | 105,671 | 0.29 | 0.08 | 405,857 | 0.08 | 0.03 |
| Kyrgyzstan | Asia | 147.29 | 528.16 | 17,736 | 0.03 | 0.01 | 405,062 | 0.48 | 0.03 |

| Country | Continent | Grid Intensity ($gCO_2e$ /kWh) | Tipping Point ($gCO_2e$ /kWh) | Available Electrification Miles (nm) | Continent Miles (%) | Global Miles (%) | Emissions Savings (kg $CO_2e$) | Continent Emissions Change (%) | Global Emissions Change (%) |
|---|---|---|---|---|---|---|---|---|---|
| Ukraine | Europe | 259.69 | 528.82 | 30,014 | 0.08 | 0.02 | 392,458 | 0.07 | 0.03 |
| Tajikistan | Asia | 116.86 | 529.05 | 17,081 | 0.03 | 0.01 | 352,437 | 0.42 | 0.03 |
| Czech Republic | Europe | 449.72 | 526.89 | 69,663 | 0.19 | 0.05 | 290,371 | 0.05 | 0.02 |
| Sudan | Africa | 263.16 | 528.69 | 14,944 | 2 | 0.01 | 196,920 | 4.19 | 0.02 |
| Romania | Europe | 240.58 | 528.08 | 11,669 | 0.03 | 0.01 | 176,701 | 0.03 | 0.02 |
| Sri Lanka | Asia | 509.78 | 528.46 | 133,591 | 0.22 | 0.1 | 128,890 | 0.15 | 0.01 |
| Norway | Europe | 30.08 | 527.34 | 4,778 | 0.01 | 0 | 117,087 | 0.02 | 0.01 |
| Russia | Asia | 441.04 | 527.25 | 18,267 | 0.03 | 0.01 | 86,797 | 0.1 | 0.01 |
| Afghanistan | Asia | 132.53 | 529.42 | 3,410 | 0.01 | 0 | 65,795 | 0.08 | 0.01 |
| Togo | Africa | 443.18 | 529.67 | 10,006 | 1.34 | 0.01 | 50,026 | 1.06 | 0 |
| Costa Rica | North America | 53.38 | 527.98 | 688 | 0.01 | 0 | 25,457 | 0.02 | 0 |
| Ghana | Africa | 484 | 529.5 | 5,475 | 0.73 | 0 | 14,268 | 0.3 | 0 |
| Uruguay | South America | 128.79 | 529.05 | 657 | 0 | 0 | 13,162 | 0 | 0 |
| Montenegro | Europe | 417.07 | 528.07 | 1,268 | 0 | 0 | 7,133 | 0 | 0 |
| Hungary | Europe | 204.19 | 527.29 | 344 | 0 | 0 | 6,207 | 0 | 0 |
| Estonia | Europe | 416.67 | 527.58 | 216 | 0 | 0 | 2,022 | 0 | 0 |
| North Macedonia | Europe | 565.35 | 527.33 | 181 | 0 | 0 | -416 | 0.04 | 0 |
| Bahamas | North America | 660.1 | 528.37 | 157 | 0 | 0 | -1,085 | 0.15 | 0 |
| Kosovo | Europe | 894.65 | 528.13 | 230 | 0 | 0 | -4,867 | 0.52 | 0 |
| Serbia | Europe | 636.06 | 528.01 | 1,656 | 0 | 0 | -8,920 | 0.95 | 0 |
| Cape Verde | Africa | 558.14 | 528.02 | 5,157 | 0.69 | 0 | -9,122 | 0.39 | 0 |
| Bosnia and Herzegovina | Europe | 600 | 528.91 | 2,637 | 0.01 | 0 | -10,119 | 1.08 | 0 |
| Azerbaijan | Europe | 671.39 | 527.15 | 2,923 | 0.01 | 0 | -25,669 | 2.74 | 0.01 |
| Dominican Republic | North America | 580.78 | 527.95 | 11,776 | 0.09 | 0.01 | -32,342 | 4.38 | 0.01 |

| Country | Continent | Grid Intensity ($gCO_2e$ /kWh) | Tipping Point ($gCO_2e$/ kWh) | Available Electrification Miles (nm) | Continent Miles (%) | Global Miles (%) | Emissions Savings (kg $CO_2e$) | Continent Emissions Change (%) | Global Emissions Change (%) |
|---|---|---|---|---|---|---|---|---|---|
| Syria | Asia | 701.66 | 529.94 | 3,445 | 0.01 | 0 | -35,267 | 0.01 | 0.01 |
| Solomon Islands | Oceania | 700 | 529.05 | 7,982 | 1.29 | 0.01 | -67,591 | 100 | 0.03 |
| Cyprus | Europe | 534.32 | 528.77 | 322,758 | 0.88 | 0.24 | -106,915 | 11.41 | 0.04 |
| Jordan | Asia | 540.92 | 527.65 | 308,305 | 0.5 | 0.22 | -222,824 | 0.09 | 0.09 |
| Egypt | Africa | 570.31 | 528.29 | 94,238 | 12.61 | 0.07 | -223,840 | 9.68 | 0.09 |
| Mauritius | Africa | 632.48 | 528.41 | 43,255 | 5.79 | 0.03 | -251,682 | 10.88 | 0.1 |
| Tunisia | Africa | 563.96 | 528.82 | 157,527 | 21.09 | 0.11 | -290,437 | 12.56 | 0.11 |
| Cuba | North America | 637.61 | 528.56 | 58,213 | 0.42 | 0.04 | -322,861 | 43.73 | 0.13 |
| Puerto Rico | North America | 678.74 | 529.21 | 39,650 | 0.29 | 0.03 | -382,002 | 51.74 | 0.15 |
| Algeria | Africa | 634.61 | 528.67 | 76,870 | 10.29 | 0.06 | -413,094 | 17.86 | 0.16 |
| Iran | Asia | 655.12 | 528.53 | 64,942 | 0.11 | 0.05 | -450,133 | 0.18 | 0.18 |
| Morocco | Africa | 630.01 | 528.08 | 95,626 | 12.8 | 0.07 | -514,790 | 22.26 | 0.2 |
| Kuwait | Asia | 649.16 | 528.83 | 93,455 | 0.15 | 0.07 | -553,934 | 0.22 | 0.22 |
| Libya | Africa | 818.69 | 529.57 | 42,193 | 5.65 | 0.03 | -609,988 | 26.37 | 0.24 |
| Oman | Asia | 564.63 | 528.49 | 425,338 | 0.69 | 0.31 | -775,011 | 0.31 | 0.3 |
| Poland | Europe | 661.93 | 528.09 | 116,668 | 0.32 | 0.09 | -779,996 | 83.25 | 0.3 |
| United Arab Emirates | Asia | 561.13 | 528.5 | 475,637 | 0.77 | 0.35 | -789,469 | 0.31 | 0.31 |
| Bangladesh | Asia | 691.41 | 526.95 | 93,884 | 0.15 | 0.07 | -831,669 | 0.33 | 0.32 |
| Lebanon | Asia | 599.01 | 528.08 | 309,113 | 0.5 | 0.23 | -1,208,790 | 0.48 | 0.47 |
| Iraq | Asia | 688.81 | 528.37 | 149,481 | 0.24 | 0.11 | -1,217,363 | 0.48 | 0.47 |
| Brunei | Asia | 893.91 | 527.42 | 58,315 | 0.09 | 0.04 | -1,253,466 | 0.5 | 0.49 |
| Israel | Asia | 582.93 | 529.02 | 400,280 | 0.65 | 0.29 | -1,263,116 | 0.5 | 0.49 |
| Thailand | Asia | 549.58 | 528.43 | 1,819,498 | 2.96 | 1.33 | -1,884,469 | 0.74 | 0.73 |
| Bahrain | Asia | 904.61 | 528.81 | 66,577 | 0.11 | 0.05 | -1,970,624 | 0.78 | 0.77 |
| Kazakhstan | Asia | 821.39 | 528.62 | 159,991 | 0.26 | 0.12 | -2,334,617 | 0.92 | 0.91 |
| Taiwan | Asia | 642.38 | 528.28 | 593,417 | 0.97 | 0.43 | -4,223,860 | 1.67 | 1.64 |

| Country | Continent | Grid Intensity ($gCO_2e$/kWh) | Tipping Point ($gCO_2e$/kWh) | Available Electrification Miles (nm) | Continent Miles (%) | Global Miles (%) | Emissions Savings (kg $CO_2e$) | Continent Emissions Change (%) | Global Emissions Change (%) |
|---|---|---|---|---|---|---|---|---|---|
| Uzbekistan | Asia | 1167.6 | 529.13 | 158,567 | 0.26 | 0.12 | -5,344,924 | 2.11 | 2.08 |
| Philippines | Asia | 610.69 | 528.64 | 3,908,976 | 6.36 | 2.85 | -16,515,928 | 6.53 | 6.42 |
| China | Asia | 582.32 | 528.17 | 6,552,850 | 10.66 | 4.78 | -18,510,915 | 7.31 | 7.2 |
| Indonesia | Asia | 675.93 | 528.79 | 2,864,686 | 4.66 | 2.09 | -21,667,281 | 8.56 | 8.43 |
| Malaysia | Asia | 605.83 | 528.2 | 8,255,728 | 13.43 | 6.02 | -32,722,020 | 12.93 | 12.72 |
| Saudi Arabia | Asia | 706.79 | 528.53 | 5,090,012 | 8.28 | 3.71 | -47,684,823 | 18.84 | 18.54 |
| India | Asia | 713.44 | 528.48 | 9,495,447 | 15.45 | 6.92 | -91,652,871 | 36.21 | 35.64 |

Table S2: Country / continent level carbon intensity and available electrification mile summary

*Section D: Country level tipping point estimation*

Here, we document our approach to estimating the tipping point for 105 countries in our model.

The tipping point grid carbon intensity for each country is calculated using the following formula:

$$CI_{grid}\ \ (gCO2e/kWh) = \frac{Total\ Fuel\ Emissions\ (gCO2)}{Total\ Energy\ for\ Electric\ Flights\ (Wh)}$$

$$= \frac{\sum_i \sum_j ((W_i \times D_j + X_i) \times CI_{fuel} \times N_j)}{\sum_i \sum_j ((Y_i \times D_j + Z_i) \times N_j)}$$

Where,

- $W_i$ represents the mass of fuel required per nautical mile travelled for aircraft model *i*.
- $X_i$ represents the mass of fuel required for takeoff for aircraft model *i*.
- $Y_i$ represents the battery energy required per nautical mile travelled for aircraft model *i*.
- $Z_i$ represents the battery energy required for takeoff for aircraft model *i*.
- $D_j$ represents the distance of route *j*.
- $N_j$ represents the number of flights on route *j*.
- $CI_{fuel}$ represents the carbon intensity of the combustion of kerosene jet fuel, 3.16 $kgCO_2/kg$.
- $CI_{grid}$ represents the carbon intensity of the electricity grid, $gCO_2e/kWh$.

For instance, in the United States, 31,918 flights are flown by the A319 for 5,131,164 nautical miles, 19,266 flights are flown by the A320 for 2,997,293 nautical miles, and 8,510 flights are flown by the A321 for 1,371,504 nautical miles. This yields 266,321,936 kg $CO_2e$ of fuel emissions, and requires 504,404,944,678 Wh of electricity, translating into a tipping point of 527.99 g $CO_2e/kWh$.

In China, 9,945 flights are flown by the A319 for 1,563,400 nautical miles, 26,743 flights are flown by the A320 for 4,261,268 nautical miles and 4,325 flights are flown by the A321 for 728,182 nautical miles. This yields 180,657,366 kg $CO_2$ of fuel emissions, and requires 342,041,630,749 Wh of electricity, translating into a tipping point of 528.17 g $CO_2e/Wh$.

Section E: *Country level rankings of emissions savings for different scenarios*

SI Table 3 provides a rank ordering of all 105 countries by, 1) overall emissions savings owing to electrification (and assuming no change in grid carbon intensity), 2) absolute change in emissions owing to a 5 percent improvement in grid carbon intensity, and 3) relative change in emissions owing to a 5 percent improvement in grid carbon intensity. Absolute and relative changes in grid intensity leverage carbon intensity estimates specified in SI Table 2.

| Country | Emissions Savings (kg CO2e) | Country | Grid Improvement (5%) Absolute Difference (kg CO2e) | Country | Grid Improvement (5%) Relative Difference (%) |
|---|---|---|---|---|---|
| Brazil | 198,824,978 | India | 17,662,915 | Cyprus | 430.03 |
| Colombia | 164,007,724 | Malaysia | 12,810,764 | Jordan | 213.56 |
| United Kingdom | 111,902,200 | China | 9,958,414 | Sri Lanka | 138.1 |
| Spain | 80,886,880 | Saudi Arabia | 9,440,817 | Thailand | 129.8 |
| United States | 79,943,985 | United States | 9,318,898 | Mexico | 119.72 |
| France | 77,224,991 | Colombia | 7,922,274 | Cape Verde | 91.75 |
| Germany | 60,529,793 | Germany | 7,839,284 | United Arab Emirates | 85.81 |
| Ecuador | 52,505,436 | Korea, Republic of | 7,835,050 | Tunisia | 79.39 |
| Greece | 40,066,534 | Vietnam | 6,866,660 | Oman | 78.03 |
| Korea, Republic of | 35,519,800 | Philippines | 6,129,279 | North Macedonia | 75.46 |
| Switzerland | 34,478,796 | Turkey | 5,677,051 | Egypt | 68.21 |
| Netherlands | 34,114,344 | Indonesia | 4,965,741 | Japan | 56.77 |
| Belgium | 20,177,007 | United Kingdom | 4,576,983 | Dominican Republic | 55.52 |
| Italy | 17,415,650 | Greece | 3,516,228 | Ghana | 54.33 |
| Turkey | 15,559,096 | Mexico | 3,372,486 | China | 53.8 |
| Vietnam | 15,193,303 | Thailand | 2,446,133 | Israel | 53.16 |
| Peru | 14,486,924 | Brazil | 2,275,802 | Vietnam | 45.2 |
| Denmark | 13,564,173 | Spain | 1,987,284 | Lebanon | 42.34 |
| Portugal | 13,240,821 | Netherlands | 1,752,591 | Bosnia and Herzegovina | 41.52 |
| Canada | 13,022,219 | Japan | 1,614,222 | Singapore | 40.84 |
| New Zealand | 12,649,638 | Cambodia | 1,495,385 | Malaysia | 39.15 |
| Austria | 9,371,865 | Italy | 1,459,973 | Philippines | 37.11 |
| Cambodia | 7,914,403 | Singapore | 1,452,936 | Turkey | 36.49 |
| Sweden | 5,433,512 | Taiwan | 1,184,775 | Malta | 33.11 |
| El Salvador | 5,002,510 | Ecuador | 1,044,337 | Morocco | 31.03 |
| Ireland | 4,121,577 | Peru | 738,009 | Mauritius | 30.18 |
| Finland | 3,625,747 | United Arab Emirates | 677,430 | Algeria | 29.88 |

| Country | Emissions Savings (kg CO2e) | Country | Grid Improvement (5%) Absolute Difference (kg CO2e) | Country | Grid Improvement (5%) Relative Difference (%) |
|---|---|---|---|---|---|
| Singapore | 3,557,615 | Israel | 671,459 | Serbia | 29.55 |
| Croatia | 2,945,408 | Oman | 604,712 | Cuba | 29 |
| Japan | 2,843,582 | Lebanon | 511,833 | Czech Republic | 28.76 |
| Mexico | 2,816,942 | Uzbekistan | 488,230 | Taiwan | 28.05 |
| Namibia | 2,792,410 | Jordan | 475,870 | Kuwait | 26.88 |
| Bulgaria | 2,283,059 | Cyprus | 459,765 | Togo | 25.96 |
| Honduras | 2,047,039 | France | 458,305 | Iran | 25.83 |
| Chile | 1,681,657 | Belgium | 357,239 | Russia | 25.42 |
| Laos | 1,640,094 | Kazakhstan | 327,245 | Pakistan | 25.08 |
| Guatemala | 1,377,309 | Canada | 309,265 | Bahamas | 24.93 |
| Nicaragua | 1,352,898 | Portugal | 302,334 | Poland | 24.78 |
| Argentina | 1,182,491 | Denmark | 273,171 | Azerbaijan | 23.38 |
| D. R. Congo | 1,067,539 | El Salvador | 264,519 | Indonesia | 22.92 |
| Venezuela | 827,215 | Iraq | 261,435 | Puerto Rico | 22.6 |
| Luxembourg | 800,850 | Ireland | 252,173 | Korea, Republic of | 22.06 |
| Bhutan | 657,577 | Bahrain | 237,053 | Iraq | 21.48 |
| Cameroon | 578,524 | Tunisia | 230,580 | Bangladesh | 21.17 |
| Pakistan | 565,444 | Bulgaria | 198,740 | Solomon Islands | 20.4 |
| Slovakia | 457,617 | Poland | 193,318 | Syria | 20.27 |
| Malta | 405,857 | Sri Lanka | 178,002 | Saudi Arabia | 19.8 |
| Kyrgyzstan | 405,062 | Bangladesh | 176,040 | India | 19.27 |
| Ukraine | 392,458 | New Zealand | 171,566 | Cambodia | 18.89 |
| Tajikistan | 352,437 | Morocco | 159,722 | Montenegro | 18.72 |
| Czech Republic | 290,371 | Brunei | 153,338 | Estonia | 18.68 |
| Sudan | 196,920 | Egypt | 152,684 | Libya | 14.1 |
| Romania | 176,701 | Kuwait | 148,904 | Kazakhstan | 14.02 |
| Sri Lanka | 128,890 | Pakistan | 141,823 | Germany | 12.95 |
| Norway | 117,087 | Malta | 134,365 | Brunei | 12.23 |

| Country | Emissions Savings (kg CO2e) | Country | Grid Improvement (5%) Absolute Difference (kg CO2e) | Country | Grid Improvement (5%) Relative Difference (%) |
|---|---|---|---|---|---|
| Russia | 86,797 | Austria | 124,427 | Kosovo | 12.19 |
| Afghanistan | 65,795 | Algeria | 123,448 | Bahrain | 12.03 |
| Togo | 50,026 | Switzerland | 121,748 | United States | 11.66 |
| Costa Rica | 25,457 | Argentina | 120,175 | Argentina | 10.16 |
| Ghana | 14,268 | Honduras | 117,496 | Uzbekistan | 9.13 |
| Uruguay | 13,162 | Iran | 116,258 | Greece | 8.78 |
| Montenegro | 7,133 | Guatemala | 113,055 | Bulgaria | 8.7 |
| Hungary | 6,207 | Chile | 103,192 | Italy | 8.38 |
| Estonia | 2,022 | Cuba | 93,632 | Guatemala | 8.21 |
| North Macedonia | -416 | Croatia | 93,442 | Cameroon | 6.86 |
| Bahamas | -1,085 | Puerto Rico | 86,344 | Chile | 6.14 |
| Kosovo | -4,867 | Libya | 86,029 | Ireland | 6.12 |
| Serbia | -8,920 | Czech Republic | 83,502 | Honduras | 5.74 |
| Cape Verde | -9,122 | Laos | 82,787 | El Salvador | 5.29 |
| Bosnia and Herzegovina | -10,119 | Mauritius | 75,968 | Netherlands | 5.14 |
| Azerbaijan | -25,669 | Nicaragua | 68,179 | Peru | 5.09 |
| Dominican Republic | -32,342 | Cameroon | 39,672 | Laos | 5.05 |
| Syria | -35,267 | Finland | 31,982 | Nicaragua | 5.04 |
| Solomon Islands | -67,591 | Sweden | 22,673 | Sudan | 4.96 |
| Cyprus | -106,915 | Venezuela | 22,453 | Ukraine | 4.83 |
| Jordan | -222,824 | Russia | 22,064 | Colombia | 4.83 |
| Egypt | -223,840 | Ukraine | 18,963 | Romania | 4.19 |
| Mauritius | -251,682 | Dominican Republic | 17,958 | United Kingdom | 4.09 |
| Tunisia | -290,437 | Namibia | 17,663 | Croatia | 3.17 |
| Cuba | -322,861 | Solomon Islands | 13,789 | Hungary | 3.15 |
| Puerto Rico | -382,002 | Togo | 12,986 | Venezuela | 2.71 |
| Algeria | -413,094 | Luxembourg | 9,976 | Spain | 2.46 |

| Country | Emissions Savings (kg CO2e) | Country | Grid Improvement (5%) Absolute Difference (kg CO2e) | Country | Grid Improvement (5%) Relative Difference (%) |
|---|---|---|---|---|---|
| Iran | -450,133 | Sudan | 9,767 | Canada | 2.37 |
| Morocco | -514,790 | Cape Verde | 8,369 | Portugal | 2.28 |
| Kuwait | -553,934 | Kyrgyzstan | 7,830 | Denmark | 2.01 |
| Libya | -609,988 | Ghana | 7,752 | Ecuador | 1.99 |
| Oman | -775,011 | Romania | 7,397 | Kyrgyzstan | 1.93 |
| Poland | -779,996 | Syria | 7,147 | Belgium | 1.77 |
| United Arab Emirates | -789,469 | Slovakia | 6,502 | Afghanistan | 1.67 |
| Bangladesh | -831,669 | Azerbaijan | 6,001 | Uruguay | 1.61 |
| Lebanon | -1,208,790 | Tajikistan | 5,003 | Slovakia | 1.42 |
| Iraq | -1,217,363 | Bosnia and Herzegovina | 4,201 | Tajikistan | 1.42 |
| Brunei | -1,253,466 | Serbia | 2,636 | New Zealand | 1.36 |
| Israel | -1,263,116 | D. R. Congo | 2,590 | Austria | 1.33 |
| Thailand | -1,884,469 | Bhutan | 1,521 | Luxembourg | 1.25 |
| Bahrain | -1,970,624 | Montenegro | 1,335 | Brazil | 1.14 |
| Kazakhstan | -2,334,617 | Afghanistan | 1,101 | Finland | 0.88 |
| Taiwan | -4,223,860 | Kosovo | 593 | Namibia | 0.63 |
| Uzbekistan | -5,344,924 | Estonia | 378 | France | 0.59 |
| Philippines | -16,515,928 | Norway | 353 | Costa Rica | 0.56 |
| China | -18,510,915 | North Macedonia | 314 | Sweden | 0.42 |
| Indonesia | -21,667,281 | Bahamas | 270 | Switzerland | 0.35 |
| Malaysia | -32,722,020 | Uruguay | 212 | Norway | 0.3 |
| Saudi Arabia | -47,684,823 | Hungary | 196 | D. R. Congo | 0.24 |
| India | -91,652,871 | Costa Rica | 143 | Bhutan | 0.23 |

Table S3: Country level rankings by total emissions savings, and absolute/relative emissions savings assuming a 5 percent improvement in grid carbon intensity

*Section F: India versus Brazil emissions comparison*

In this section, we scrutinize the factors that explain emissions differences observed between India and Brazil. Compared to fuel emissions, electrifying flights in India results in an increase of 91,652,871 kg $CO_2$e, while electrifying flights in Brazil results in a decrease of 198,824,978 kg $CO_2$e. The following analysis determines how much of this absolute difference is accounted for by the grid, varied number of flights (which influences takeoff emissions), varied distance travelled and varied aircraft composition, i.e. the proportion of flights served by different aircraft.

We express the absolute increase or decrease in emissions from electrification for a country as follows:

$$Fuel\ Emissions\ (kgCO2) = \sum_i ((W_i \times F_i \times Total\ Flights) + (X_i \times G_i \times Total\ Distance)) \times CI_{fuel})$$

$$Electricity\ Emissions\ (kgCO2e) = \sum_i ((Y_i \times F_i \times Total\ Flights + (Z_i \times G_i \times Total\ Distance)) \times CI_{grid}$$

$$Absolute\ Difference\ (gCO2e) = |Fuel\ Emissions - Electricity\ Emissions|$$

Where,

- $F_i$ represents the fraction of total flights flown by a certain aircraft.
- $G_i$ represents the fraction of total miles travelled by a certain aircraft.

All other variables are the same as defined above.

India's grid carbon intensity (713.44 gCO2e/kWh) is 35 percent higher than its tipping point of 528.48 gCO2e/kWh. Adjusting $CI_{grid,}$ for Brazil to an intensity 35 percent *lower* than India's tipping point, 343.52 $gCO_2e/kWh$, emissions savings fall to 85,358,311 kg $CO_2$e, lower than our 'target' of 91,652,871 kg $CO_2$e. Adjusting $Total\ Flights$ in Brazil to match that of India (while preserving Brazil's original parameters for $F_i, G_i, Total\ Distance$) increases emissions savings to 90,755,509 kg $CO_2$e. Increasing the distance travelled for flights originating from Brazil to match that of India (while preserving Brazil's original parameters for $F_i, G_i$) increases emissions savings to 92,799,000 kg $CO_2$e. The remaining absolute difference of 1,146,129 kg $CO_2$e is accounted for by the difference in aircraft composition for flights.

Discerning the difference in the absolute value of the increase in carbon emissions and emissions savings in India and Brazil (107,172,107 kg $CO_2$e), 106 percent may be attributed to the difference in grid intensity (113,466,667 kg $CO_2$e), -5 percent may be attributed to the difference in the number of flights and thus takeoffs (-5,397,198 kg $CO_2$e), -2 percent may be attributed to the difference in the total number of nautical miles travelled (-2,043,491 kg $CO_2$e), and the remaining 1 percent is accounted for by differences in aircraft composition (1,146,129 kg $CO_2$e)[4].

[4] A positive percentage 'reduces' the gap between carbon emissions avoided in India and carbon emissions saved in Brazil, and a negative percentage 'increases' the gap.

*Section G: Sensitivity analysis*

Table S4 provides results from a exceedance sensitivity analysis based on 5 percent change in key model parameters. These parameters are passenger weight (5 percent decrease), thermal efficiency of fossil fuel powered engine (5 percent decrease), thermal efficiency of battery electric engine (5 percent increase), energy density (5 percent increase), and MLW (5 percent increase).

| Aircraft Model | Current Exceedance | Passenger Weight (kg) | Thermal Efficiency (Fossil Fuel) (%) | Thermal Efficiency (Battery Electric) (%) | Energy Density (Wh/kg) | Max. Landing Weight (kg) |
|---|---|---|---|---|---|---|
| EMB-120 | 1.81<br>0 | 1.8<br>-0.7 | 1.77<br>-2.57 | 1.77<br>-2.45 | 1.77<br>-2.45 | 1.73<br>-4.76 |
| Dornier 328 | 2.19<br>0 | 2.17<br>-0.5 | 2.12<br>-3 | 2.12<br>-2.86 | 2.12<br>-2.86 | 2.08<br>-4.76 |
| EMB135 | 1.61<br>0 | 1.6<br>-0.59 | 1.57<br>-2.51 | 1.57<br>-2.39 | 1.57<br>-2.39 | 1.53<br>-4.76 |
| DHC-8 | 1.42<br>0 | 1.41<br>-0.77 | 1.39<br>-2.15 | 1.39<br>-2.05 | 1.39<br>-2.05 | 1.35<br>-4.76 |
| ATR 42 | 1.69<br>0 | 1.67<br>-0.82 | 1.65<br>-2.32 | 1.65<br>-2.21 | 1.65<br>-2.21 | 1.61<br>-4.76 |
| EMB145 | 1.66<br>0 | 1.65<br>-0.67 | 1.62<br>-2.38 | 1.62<br>-2.27 | 1.62<br>-2.27 | 1.58<br>-4.76 |
| CRJ200 | 1.83<br>0 | 1.81<br>-0.67 | 1.78<br>-2.58 | 1.78<br>-2.46 | 1.78<br>-2.46 | 1.74<br>-4.76 |
| ATR 72 | 1.57<br>0 | 1.56<br>-0.98 | 1.54<br>-2.13 | 1.54<br>-2.03 | 1.54<br>-2.03 | 1.5<br>-4.76 |
| CRJ700 | 1.7<br>0 | 1.69<br>-0.67 | 1.66<br>-2.42 | 1.66<br>-2.3 | 1.66<br>-2.3 | 1.62<br>-4.76 |
| EMB170 | 1.52<br>0 | 1.51<br>-0.76 | 1.49<br>-2.18 | 1.49<br>-2.07 | 1.49<br>-2.07 | 1.45<br>-4.76 |
| EMB175 | 1.5<br>0 | 1.49<br>-0.77 | 1.47<br>-2.08 | 1.47<br>-1.98 | 1.47<br>-1.98 | 1.43<br>-4.76 |
| CRJ900 | 1.86<br>0 | 1.84<br>-0.69 | 1.81<br>-2.58 | 1.81<br>-2.45 | 1.81<br>-2.45 | 1.77<br>-4.76 |
| RJ85 | 1.71<br>0 | 1.7<br>-0.74 | 1.67<br>-2.36 | 1.67<br>-2.25 | 1.67<br>-2.25 | 1.63<br>-4.76 |
| CRJ1000 | 1.53<br>0 | 1.52<br>-0.84 | 1.5<br>-2.12 | 1.5<br>-2.02 | 1.5<br>-2.02 | 1.46<br>-4.76 |
| EMB190 | 1.5<br>0 | 1.49<br>-0.73 | 1.47<br>-2.06 | 1.47<br>-1.97 | 1.47<br>-1.97 | 1.43<br>-4.76 |
| B717-200 | 1.54<br>0 | 1.53<br>-0.76 | 1.51<br>-2.13 | 1.51<br>-2.03 | 1.51<br>-2.03 | 1.47<br>-4.76 |
| A318-100 | 1.34<br>0 | 1.33<br>-0.7 | 1.31<br>-1.86 | 1.31<br>-1.77 | 1.31<br>-1.77 | 1.27<br>-4.76 |
| B737-500 | 1.58<br>0 | 1.57<br>-0.69 | 1.55<br>-2.33 | 1.55<br>-2.22 | 1.55<br>-2.22 | 1.51<br>-4.76 |

| Aircraft Model | Current Exceedance | Passenger Weight (kg) | Thermal Efficiency (Fossil Fuel) (%) | Thermal Efficiency (Battery Electric) (%) | Energy Density (Wh/kg) | Max. Landing Weight (kg) |
|---|---|---|---|---|---|---|
| EMB195 | 1.57<br>0 | 1.56<br>-0.77 | 1.54<br>-2.19 | 1.54<br>-2.09 | 1.54<br>-2.09 | 1.5<br>-4.76 |
| Fokker 100 | 1.71<br>0 | 1.69<br>-0.82 | 1.67<br>-2.37 | 1.67<br>-2.26 | 1.67<br>-2.26 | 1.63<br>-4.76 |
| B737-600 | 1.57<br>0 | 1.56<br>-0.65 | 1.54<br>-2.25 | 1.54<br>-2.14 | 1.54<br>-2.14 | 1.5<br>-4.76 |
| B737-300 | 1.6<br>0 | 1.59<br>-0.74 | 1.56<br>-2.3 | 1.56<br>-2.19 | 1.56<br>-2.19 | 1.52<br>-4.76 |
| B737-700 | 1.52<br>0 | 1.51<br>-0.73 | 1.49<br>-2.16 | 1.49<br>-2.06 | 1.49<br>-2.06 | 1.45<br>-4.76 |
| A319-100 | 1.3<br>0 | 1.29<br>-0.81 | 1.28<br>-2.02 | 1.28<br>-1.92 | 1.28<br>-1.92 | 1.24<br>-4.76 |
| MD 80 | 1.46<br>0 | 1.45<br>-0.84 | 1.43<br>-2.11 | 1.43<br>-2.01 | 1.43<br>-2.01 | 1.39<br>-4.76 |
| B737-400 | 1.56<br>0 | 1.55<br>-0.82 | 1.53<br>-2.26 | 1.53<br>-2.15 | 1.53<br>-2.15 | 1.49<br>-4.76 |
| A320-200 | 1.3<br>0 | 1.29<br>-0.91 | 1.28<br>-1.91 | 1.28<br>-1.82 | 1.28<br>-1.82 | 1.24<br>-4.76 |
| B737-800 | 1.5<br>0 | 1.49<br>-0.84 | 1.47<br>-2.04 | 1.47<br>-1.94 | 1.47<br>-1.94 | 1.43<br>-4.76 |
| B737-900 | 1.52<br>0 | 1.51<br>-0.86 | 1.49<br>-2.03 | 1.49<br>-1.93 | 1.49<br>-1.93 | 1.45<br>-4.76 |
| B757-200 | 1.4<br>0 | 1.39<br>-0.7 | 1.38<br>-1.95 | 1.38<br>-1.85 | 1.38<br>-1.85 | 1.34<br>-4.76 |
| A321-200 | 1.37<br>0 | 1.35<br>-0.9 | 1.34<br>-1.89 | 1.34<br>-1.8 | 1.34<br>-1.8 | 1.3<br>-4.76 |
| B767-200 | 1.3<br>0 | 1.29<br>-0.62 | 1.27<br>-1.96 | 1.27<br>-1.87 | 1.27<br>-1.87 | 1.24<br>-4.76 |
| B787-8 | 1.21<br>0 | 1.2<br>-0.55 | 1.19<br>-1.76 | 1.19<br>-1.68 | 1.19<br>-1.68 | 1.15<br>-4.76 |
| B767-300 | 1.22<br>0 | 1.21<br>-0.7 | 1.2<br>-1.8 | 1.2<br>-1.71 | 1.2<br>-1.71 | 1.16<br>-4.76 |
| B767-400 | 1.28<br>0 | 1.27<br>-0.63 | 1.26<br>-1.84 | 1.26<br>-1.75 | 1.26<br>-1.75 | 1.22<br>-4.76 |
| A330-200 | 1.25<br>0 | 1.24<br>-0.57 | 1.23<br>-1.82 | 1.23<br>-1.74 | 1.23<br>-1.74 | 1.19<br>-4.76 |

| Aircraft Model | Current Exceedance | Passenger Weight (kg) | Thermal Efficiency (Fossil Fuel) (%) | Thermal Efficiency (Battery Electric) (%) | Energy Density (Wh/kg) | Max. Landing Weight (kg) |
|---|---|---|---|---|---|---|
| B787-9 | 1.15<br>0 | 1.14<br>-0.62 | 1.13<br>-1.89 | 1.13<br>-1.8 | 1.13<br>-1.8 | 1.1<br>-4.76 |
| A340-300 | 1.32<br>0 | 1.32<br>-0.55 | 1.3<br>-1.87 | 1.3<br>-1.78 | 1.3<br>-1.78 | 1.26<br>-4.76 |
| A340-500 | 1.32<br>0 | 1.32<br>-0.47 | 1.3<br>-1.89 | 1.3<br>-1.8 | 1.3<br>-1.8 | 1.26<br>-4.76 |
| A330-300 | 1.29<br>0 | 1.28<br>-0.63 | 1.27<br>-1.79 | 1.27<br>-1.7 | 1.27<br>-1.7 | 1.23<br>-4.76 |
| A350-900 | 1.26<br>0 | 1.25<br>-0.59 | 1.24<br>-1.67 | 1.24<br>-1.59 | 1.24<br>-1.59 | 1.2<br>-4.76 |
| B777-200 | 1.39<br>0 | 1.38<br>-0.57 | 1.36<br>-1.99 | 1.37<br>-1.89 | 1.37<br>-1.89 | 1.33<br>-4.76 |
| A340-600 | 1.17<br>0 | 1.16<br>-0.6 | 1.15<br>-1.52 | 1.15<br>-1.45 | 1.15<br>-1.45 | 1.11<br>-4.76 |
| A350-1000 | 1.24<br>0 | 1.23<br>-0.62 | 1.22<br>-1.74 | 1.22<br>-1.65 | 1.22<br>-1.65 | 1.18<br>-4.76 |
| B747-400 | 1.33<br>0 | 1.32<br>-0.57 | 1.3<br>-2.08 | 1.3<br>-1.98 | 1.3<br>-1.98 | 1.26<br>-4.76 |
| A380-800 | 1.61<br>0 | 1.61<br>-0.39 | 1.57<br>-2.42 | 1.57<br>-2.3 | 1.57<br>-2.3 | 1.53<br>-4.76 |
| B777-300 | 1.47<br>0 | 1.46<br>-0.75 | 1.44<br>-1.97 | 1.44<br>-1.88 | 1.44<br>-1.88 | 1.4<br>-4.76 |

Table S4: Exceedance sensitivity analysis for key model parameters. Top line represents absolute exceedance change and bottom line reflects relative exceedance change